\documentclass[a4paper,11pt]{article}
\pdfoutput=1
\usepackage{jcappub}

\newcommand\pos[1]{\href{https://pos.sissa.it/contribution?id=#1}{\tt #1}}
\newcommand{\arXivid}[1]{\href{https://arxiv.org/abs/#1}{\tt arXiv:#1}}

\newcommand{\hepph}[1]{\href{https://arxiv.org/abs/hep-ph/#1}{\tt hep-ph/#1}}

\newcommand{\astroph}[1]{\href{https://arxiv.org/abs/astro-ph/#1}{\tt astro-ph/#1}}

\newcommand{\Math}[2]{%
\if!#1!%
\href{https://arxiv.org/abs/math/#2}{\tt math/#2}%
\else%
\href{https://arxiv.org/abs/math.#1/#2}{\tt math.#1/#2}%
\fi}
\newcommand\inspire[1]{\href{https://inspirehep.net/search?p=find+#1}{{\tiny IN}{\footnotesize SPIRE}}}
\newcommand{\jcap}[3] {\href{https://doi.org/10.1088/1475-7516/#2/#1/#3}{\emph{JCAP} {\bf #1} (#2) #3}}
\newcommand{\jhep}[3] {\ifnum#2>2009%
\href{https://doi.org/10.1007/JHEP#1(#2)#3}{\emph{JHEP} {\bf #1} (#2) #3}%
\else%
\href{https://doi.org/10.1088/1126-6708/#2/#1/#3}{\emph{JHEP} {\bf #1} (#2) #3}%
\fi}
\newcommand\erratum[4][ibid.\ ]{\emph{Erratum #1}{\bf #2} (#3) #4}

\usepackage{rotating}
\usepackage{bm}

\newcommand{\be}{\begin{equation}}
\newcommand{\ee}{\end{equation}}

\newcommand{\GeV}{{\rm GeV}}

\newcommand{\kpc}{{\rm kpc}}

\newcommand{\cm}{{\rm cm}}

\newcommand{\dm}{{\textrm{\tiny DM}}}

\title{On velocity-dependent dark matter annihilations in dwarf satellites}

\author[a,b]{Mihael Peta\v{c},}
\author[a,b]{Piero Ullio}
\author[c]{and Mauro Valli}

\affiliation[a]{SISSA,\\
Via Bonomea, 265, 34136 Trieste, Italy}
\affiliation[b]{INFN --- Sezione di Trieste,\\
Via Bonomea 265, 34136 Trieste, Italy}
\affiliation[c]{INFN --- Sezione di Roma,\\
P.le A. Moro 2, I-00185 Roma, Italy}

\emailAdd{mihael.petac@sissa.it}
\emailAdd{ullio@sissa.it}
\emailAdd{mauro.valli@roma1.infn.it}

\abstract{Milky Way dwarf spheroidal satellites are a prime target for Dark  Matter (DM) indirect searches. Recently the importance of possible long-range interactions has been recognized, as they can boost the expected DM gamma ray signal by orders of magnitude through an effect commonly known as the Sommerfeld enhancement. However, for such analyses precise modelling of DM phase-space distribution becomes crucial and can introduce large uncertainties in the final result. We provide a pioneering attempt towards a comprehensive investigation of these systematics. First, the DM halo profiles are constrained using Bayesian inference on the available stellar kinematic datasets with a careful treatment of observational and theoretical uncertainties. We consider both cuspy and cored parametric DM density profiles, together with the case of a non-parametric halo modelling directly connected to observable quantities along the line-of-sight. After reconsidering the study case of ergodic systems, the basic ingredient of all previous analyses, we investigate for the first time scenarios where DM particles are allowed to have anisotropic velocity distributions. Referring to a generalized $J$-factor, sensitive to velocity-dependent effects, an enhancement (suppression) with respect to the isotropic phase-space distributions is obtained for the case of tangentially (radially) biased DM particle orbits. We provide new estimates for $J$-factors for the eight brightest Milky Way dwarfs also in the limit  of velocity-independent DM annihilation, in good agreement with previous results in literature, and derive data-driven lower-bounds based on the non-parametric modelling of the halo density. This work presents a state-of-the-art analysis of the aforementioned effects and falls within the interest of current and future experimental collaborations involved in DM indirect detection programs.}

\begin{document}
\maketitle\flushbottom

\section{Introduction}

Dwarf spheroidal (dSph) satellites in the Milky Way (MW) are a unique laboratory for the investigation --- and most importantly for possible detection --- of signals from particle dark matter (DM)~\cite{Lake-ml-1990du,Evans-ml-2003sc}. This follows form the fact that they have fairly large DM densities and small amount of contaminants from standard astrophysical sources: some of the largest known mass-to-light ratios~\cite{bib2002MNRAS.333..697L,Walker2009}, typically M/L~$\gtrsim \mathcal{O}(10^{2})$~M$_{\odot}$/L$_{\odot}$, are indeed found for this class of objects; at the same time, MW  dSphs show no sign of recent star formation activity and negligible emissivities associated to their interstellar medium~\cite{bib2012AJ_McConnachie,WalkerReviewMWdSphs,Battaglia-ml-2013wqa}.

In case DM particles can annihilate in pairs, like e.g. Weakly Interacting Massive Particles (WIMPs)~\cite{PhysRevLett.39.165}, being a thermal relics from the early Universe, the prompt emission of photons is one of the most important DM indirect detection signatures~\cite{bib2014arXiv1411.1925C,bib2017arXiv171005137S}. In particular, for masses at the GeV scale or above, promising signal-to-noise ratios are expected in the gamma-ray band when observing MW dSphs~\cite{Walker-ml-2011fs,Charbonnier-ml-2011ft,Martinez-ml-2013els,Geringer-Sameth-ml-2014yza,Bonnivard-ml-2015xpq}, making it possible to derive relevant constraints on the fundamental properties of the DM particle~\cite{Ackermann-ml-2011wa,Ackermann-ml-2013yva,Ackermann-ml-2015zua,Ahnen-ml-2016qkx,Fermi-LAT-ml-2016uux}.

Considering a given object whose DM phase-space distribution function (PSDF) is $f_\dm(\vec{x}, \vec{v})$ (the PSDF tracks the mass density per phase-space element at a given point in space $\vec{x}$), the gamma-ray flux due to pair annihilation of DM particles in a telescope pointing at the object with angular acceptance $\Delta \Omega$ would be:
\begin{equation}
\label{eq:gen_annihilation_flux}
\frac{d \Phi_{\gamma}}{dE_{\gamma}} = \frac{1}{4 \pi} \frac{(\overline{\sigma v}_{\mathrm{rel}})}{2 m^2_{\chi}} \frac{dN_{\gamma}}{dE_{\gamma}} \, \int_{\Delta \Omega} d \Omega \int_{\textrm{l.o.s.}} d \ell \int d \vec{v}_1 f_{\textrm{\tiny DM}} (\vec{x}, \vec{v}_1) \int d \vec{v}_2 \;  f_{\textrm{\tiny DM}}(\vec{x}, \vec{v}_2) \, S( | \vec{v}_{\textrm{rel}}|) \,,
\end{equation}
where we have assumed that the DM particle $\chi$ is its own antiparticle (otherwise an extra factor of 1/2 is needed), as well as defined $m_{\chi}$ to be its mass and $dN_{\gamma}/dE_{\gamma}$ the energy spectrum of the photons produced per annihilation. In this formula we are paying attention to cases in which the pair annihilation cross section $({\sigma v}_{\mathrm{rel}})$ has a non-trivial dependence on the modulus of the relative velocity $|\vec{v}_\textrm{rel}| = |\vec{v}_1 - \vec{v}_2|$, with $\vec{v}_1$ and $\vec{v}_2$ being the velocities of two annihilating particles; we have then factorized $({\sigma v}_{\mathrm{rel}})$ into the velocity independent term $(\overline{\sigma v}_{\mathrm{rel}})$ times a dimensionless factor fully comprising its dependence on relative velocity, $({\sigma v}_{\mathrm{rel}}) = (\overline{\sigma v}_{\mathrm{rel}}) \cdot S( | \vec{v}_{\textrm{rel}}|)$. The gamma-ray flux is then obtained by folding $S$ onto the two velocity distributions and integrating over all contributions along the l.o.s. and within the instrument's angular acceptance; we isolate these last two steps introducing:
\begin{align}
\label{eq:j_factor}
J \equiv & \int_{\Delta \Omega} d \Omega \int_{\textrm{l.o.s.}} d \ell \int d \vec{v}_1 f_{\textrm{\tiny DM}} (\vec{x}, \vec{v}_1) \int d \vec{v}_2 \;  f_{\textrm{\tiny DM}}(\vec{x}, \vec{v}_2) \, S( | \vec{v}_{\textrm{rel}}|) \nonumber\\
\equiv & \int_{\Delta\Omega} d\Omega \int_{\rm l.o.s.} d\ell \, \rho_{\textrm{ \tiny DM}}^2(\vec{x}) \ \langle S (v_{\textrm{rel}}) \rangle (\vec{x}) \,,
\end{align}
where this definition is in analogy to what is usually denoted in the literature as ``$J$-factor'', focussing on the standard lore with S-wave annihilation of non-relativistic particles, in which $({\sigma v}_{\mathrm{rel}})$ is, to a good approximation, velocity independent (we have normalized $S (| \vec{v}_{\textrm{rel}}|)$ to 1 in such a case).

There are several scenarios in which a non-trivial $S (| \vec{v}_{\textrm{rel}}|)$ arises; e.g., there are models for which S-wave annihilations are forbidden or severely suppressed, and hence P-wave processes become relevant, inducing a $| \vec{v}_{\textrm{rel}}|^2$ scaling.
In this paper we focus on cases in which non-perturbative effects due to long range interactions introduce an additional velocity dependence proportional to inverse powers of $v_{\textrm{rel}}$; since DM particles in galactic halos are non-relativistic, this leads to a large increase in the expected fluxes for indirect searches~\cite{Hisano-ml-2003ec,Hisano-ml-2004ds,Pospelov-ml-2007mp,ArkaniHamed-ml-2008qn,Feng-ml-2009hw,Shepherd-ml-2009sa, vonHarling-ml-2014kha,Ovanesyan-ml-2016vkk}. The effect, commonly known as Sommerfeld enhancement, is especially important for small objects, like dSphs, where DM particles are expected to have fairly small velocities. The fact that accurate predictions for the gamma-ray flux induced by DM annihilations require a careful study of the phase-space distribution $f_\dm(\vec{x}, \vec{v})$  --- as apposed to results approximating it with velocity moments --- was acknowledged only recently, see e.g.~\cite{Boddy-ml-2017vpe,Lu-ml-2017jrh,Bergstrom-ml-2017ptx}, with the notable exception of the pioneering study in ref.~\cite{Ferrer-ml-2013cla}.

Our paper is devoted to a detail study of the Sommerfeld enhancement effect in MW dSphs. We address uncertainties related to possible choices of the DM phase-space distribution function, considering a few alternative scenarios, including cases with anisotropic DM velocity distributions, which are analyzed for the first time in this context (see also the recent analysis~\cite{bib2018JCAP...09..040L}, which appeared during the peer review process of this work). Furthermore, the PSDF models considered here are fully consistent with the corresponding DM density profiles to be fitted to dSph observational data; we discuss results for two different parametric forms for the density profile, the NFW profile, which is singular towards the center of the system, and the Burkert profile, which instead has a flat core, as well as check predictions for a non-parametric profile obtained from an inversion of the Jeans equation for spherical systems. A growing effort that has been dedicated in recent years to the characterization of the DM content in these objects, see, e.g.,~\cite{Strigari-ml-2008ib,Walker2009,Wolf-ml-2009tu,Hayashi-ml-2012si}; we apply here a specular techniques to efficiently scan our large --- up to seven dimensional --- parameter spaces.
Using Bayesian inference we compute the standard, as well as Sommerfeld-enhanced, $J$-factors for all eight so-called ``classical'' dSphs (that were already known before the first discoveries of ultra-faint ones), for which the quality of kinematic data is adequate to provide fairly small statistical errors. We demonstrate that the DM velocity distribution --- about which we have virtually no information --- induces significant uncertainties in models with velocity dependent DM annihilation cross-section.

The paper is structured as follows: in section~\ref{sec:theory_mass} we describe the theoretical background and introduce the different DM PSDFs, which are considered in the current analysis. Section~\ref{sec:bayesian} contains our main results, including the astrophysical $J$-factors in different regimes of Sommerfeld enhancement, as obtained through Bayesian analysis of recent dSph data. Finally, section~\ref{sec:concl} contains our conclusions.


\section{Phase-space density functions: basics and beyond}
\label{sec:theory_mass}

In this section we present the basic theoretical elements  of our study.
We introduce the definition of PSDF and other quantities that are relevant for the present analysis. In particular, we devote our attention to various prescriptions for reconstructing the phase-space distribution of DM particles, allowing for standard isotropic as well as anisotropic models, which were left unaddressed by previous works.

The PSDF may be regarded as one of the key concepts in the theoretical study of galactic dynamics.
Both DM particle and stellar ensembles can be indeed specified by their distribution functions,  $f_{\textrm{\tiny DM}}(t,\vec{x},\vec{v})$ and $f_{\star}(t,\vec{x},\vec{v})$ respectively (to shorten the notation, below we denote by $q_{\textrm{\tiny DM}, \star}$ a given quantity $q$ referring either to the DM or to the stellar component; in all equations there is no mixing between the two cases). In the steady-state limit, their 0-$th$ moment corresponds to the density profile:
\begin{equation}
\label{eq:density}
\rho_{\textrm{\tiny DM}, \star}(\vec{x}) = \int d^3 v \; f_{\textrm{\tiny DM}, \star}(\vec{x}, \vec{v}) \,,
\end{equation}
while first and second moments carry information about mean velocities and velocity dispersions in the system according to:
\begin{eqnarray}
\label{eq:velocity_Jeans}
 \bar{v}_{i\, \textrm{\tiny DM},\star}(\vec{x}) & =  & \frac{1}{\rho_{\textrm{\tiny DM},\star}(\vec{x})} \int d^3 v \; v_{i}\; f_{\textrm{\tiny DM}, \star}(\vec{x}, \vec{v}) \,, \nonumber \\
\sigma^2_{ij\, \textrm{\tiny DM},\star} (\vec{x})  & =  & \frac{1}{\rho_{\textrm{\tiny DM},\star}(\vec{x})} \int d^3 v \; v_{i}v_{j} \; f_{\textrm{\tiny DM}, \star}(\vec{x}, \vec{v}) -\bar{v}_{i\, \textrm{\tiny DM},\star}(\vec{x})\bar{v}_{j\, \textrm{\tiny DM},\star}(\vec{x}) \,,
\end{eqnarray}
with $i$ and $j$ labelling the three components of the velocity.
Note that eqs.~\eqref{eq:density}--\eqref{eq:velocity_Jeans} allow us to undertake a statistical description of the evolution of both, DM particles and stars, as collisionless relaxed systems. Starting from the
appropriate Boltzmann equation for the PSDF and properly combining its 0-$th$ and first moments, we can derive the well-known Jeans equations~\cite{bt08}. Taking the steady-state limit as a valid regime for the whole system, the Jeans equations in spherical approximation collapse into a single differential equation:
\begin{equation}
\label{eq:SphJeansEq}
\frac{d p_{r \, \textrm{\tiny DM},\star}}{d r} + \frac{2 \beta_{\textrm{\tiny DM},\star} (r)}{r} p_{r \, \textrm{\tiny DM},\star}(r) = - \rho_{\textrm{\tiny DM},\star}(r) \frac{d \Phi}{dr}.
\end{equation}
Here $p_{r\, \textrm{\tiny DM},\star} \equiv \rho_{\, \textrm{\tiny DM},\star} \, \sigma^{2}_{rr\, \textrm{\tiny DM},\star} $ correspond to the radial dynamical pressure of DM particles and stars; $\Phi$ constitutes the total gravitational potential; $\beta_{\textrm{\tiny DM}, \star}$ are the DM and stellar orbital anisotropies:
\begin{equation}
\label{eq:beta_def}
\beta_{\dm,\star}(r) = 1 - \frac{\sigma^2_{\varphi \varphi\, \textrm{\tiny DM},\star} + \sigma^2_{\theta \theta\, \textrm{\tiny DM},\star}}{2 \sigma^2_{rr\, \textrm{\tiny DM},\star}} \,,
\end{equation}
measuring the departure from the isotropic regime, represented by $\beta_{\dm, \star} = 0$, while in the case of pure radial (circular) orbits they take the value of 1 ($-\infty)$.

The spherical Jeans equation for the stellar component provides a connection between the stellar tracers and the underlying gravitational potential, which in MW dSph satellites can be safely assumed to be due to the DM component only. In section~\ref{sec:bayesian} we use it as a tool to constrain the DM halo profiles. In particular we will consider two approaches, one in which we assume a parametric functional form for $\rho_\dm(r)$ and fit the relative parameters to the kinematic observables, and another where we directly use the measured stellar velocity dispersion to ``invert'' eq.~\eqref{eq:SphJeansEq} and derive an explicit expression for the DM density profile. Regarding the parametric profiles,
we will consider two models commonly used in the literature, the NFW profile~\cite{Navarro-ml-1996gj}, motivated by results of N-body simulations of hierarchical clustering in $\Lambda$CDM cosmologies:
\begin{equation}
\label{eq:rho_NFW}
\rho_{\textrm{\tiny NFW}}  =  \frac{\rho_{s}}{(r/r_{s})\left(1+r/r_{s}\right)^{2}}  \,,
\end{equation}
and the Burkert profile~\cite{Burkert-ml-1995yz}, namely a phenomenological model with an inner constant density rather than the cuspy behavior predicted by pure
cold DM simulations:
\begin{equation}
\label{eq:rho_BUR}
\rho_{\textrm{\tiny BUR}}  =  \frac{\rho_{s}}{\left(1+r/r_{s}\right)\left(1+(r/r_{s})^{2}\right)} \ .
\end{equation}
The details for the non-parametric model are given in refs.~\cite{Ullio-ml-2016kvy,Valli-ml-2016xtu}. In summary, the DM density profile is fully specified in terms of  two quantities that can be fitted to observational data, namely the stellar surface density and the l.o.s. projected velocity dispersion (involving essentially a l.o.s. projection onto the observational plane of $\rho_{\star}$ and $p_{r\,\star}$), plus another quantity, the stellar anisotropy profile $\beta_{\star}(r)$, that needs to be selected according to some prior
(or prejudice, since there is no observational handle and only weak theoretical guidance on this quantity). Applying this approach in full generality to our current analysis is computationally challenging. In this study we will limit ourself to the sample case of $\beta_{\star}(r) \rightarrow - \infty$, and the approximation of constant l.o.s. projected velocity dispersion and cored stellar profile, as supported by spectroscopic and photometric data for the classical dSphs; it was shown in ref.~\cite{Ullio-ml-2016kvy} that this corresponds the lowest possible $J$-factor in the limit of velocity-independent DM annihilation rates. Once the density profile has been successfully reconstructed, we then proceed building self-consistent extrapolations regarding the PSDF.

\subsection{Maxwell-Boltzmann velocity profile}

Most analyses concerning Sommerfeld enhancement of indirect detection signals have assumed velocity distributions which are of Maxwell-Boltzmann type, as partially motived by the \emph{isothermal sphere} model. The latter is defined by a constant velocity dispersion and results in Maxwellian velocity distribution at all radii. While the assumption of constant velocity dispersion is in general too crude for this kind of analysis and corresponds only to the radial profile of a \emph{singular isothermal sphere}, one can readjust it to a dynamical model at hand by applying the spherical Jean's equation for the DM component in the limit of $\beta_\dm(r) = 0$, finding the radially-dependent velocity dispersion expressed by the formula:
\begin{equation}
\label{eq:MB_velocity_dispersion}
\sigma_{\textrm{\tiny DM}}^2(r) = \frac{1}{\rho_{\textrm{\tiny DM}}(r)} \int_r^{\infty} dr' \rho_{\textrm{\tiny DM}}(r') \frac{d\Phi}{d r'} \, .
\end{equation}
One can then approximate the PSDF by:
\begin{equation}
\label{eq:MB_PSDF}
f_{\textrm{\tiny DM-MB}}(r, v) = \frac{\rho_{\textrm{\tiny DM}}(r)}{(2 \pi \sigma_\dm^2(r))^{3/2}} \cdot \exp\left[ -v^{2}/ \left(2 \sigma^{2}_{\textrm{\tiny DM}}(r)\right)\right] \,,
\end{equation}
where $v$ is the modulus of the (isotropic) velocity.

\subsection{PSDF from Eddington's inversion formula}

According to the Jeans theorem any steady state solution of the Boltzmann equation can be parametrized in terms of integrals of motion. This is particularly simple for spherical isotropic systems, in which the PSDF depends only on the particle's energy. It is customary to express this dependence in terms of the relative energy:
\begin{align}
	\mathcal{E} \equiv \Psi(r) - \frac{v^2}{2} \,,
\end{align}
where we have introduced the relative potential $\Psi(r) \equiv \Phi_{b}-\Phi(r)$. Since $\Psi(r)$ is a monotonic function, one can treat $\rho_\dm$ as a function of $\Psi$ and invert the expression in eq.~\eqref{eq:density} to find the so-called \emph{Eddington's inversion formula}~\cite{bt08}:
\begin{align}
\label{eq:Eddington}
f_{\textrm{\tiny DM-E}}(\mathcal{E}) =\; &  \frac{1}{\sqrt{8} \pi^2} \frac{d}{d \mathcal{E}} \int_{0}^{\mathcal{E}} \frac{d \Psi}{\sqrt{\mathcal{E} - \Psi}} \frac{d \rho_{\textrm{\tiny DM}}}{d \Psi} \nonumber \\
=\; & \frac{1}{\sqrt{8} \pi^2} \left[ \int_{0}^{\mathcal{E}} \frac{d \Psi}{\sqrt{\mathcal{E} - \Psi}} \frac{d^2 \rho_{\textrm{\tiny DM}}}{d \Psi^2} - \left. \left( \frac{1}{\sqrt{\mathcal{E} - \Psi}} \frac{d \rho_{\textrm{\tiny DM}}}{d \Psi} \right) \right\rvert_{\Psi = 0} \right] \, .
\end{align}
The choice of $\Phi_b$ must be such that for any $\mathcal{E} < 0$ the PSDF vanishes. We consider the case of density profiles that are truncated at a given maximum radius $r_b$, imposing $\Psi(r_b) = 0$ (see appendix~\ref{app:boundary} for details).
PSDFs computed through Eddington's inversion generally deviate from the Maxwell-Boltzmann approximation, constructed according to eq.~\eqref{eq:MB_PSDF}--\eqref{eq:MB_velocity_dispersion}, with the difference being most significant for cuspy DM density profiles.

\subsection{Anisotropic PSDFs}

Going beyond the standard lore, we also consider spherical models which do not rely on the assumption of isotropic DM velocity distribution. In fact, similar approaches to Eddington's inversion allow to reconstruct DM phase-space distribution with non-vanishing anisotropy profiles~\cite{bt08}.

For spherically symmetric systems, Jean's theorem implies that the PSDFs can be parametrized in terms of two integrals of motion, namely the relative energy $\mathcal{E}$ and the magnitude of angular momentum $L$, i.e.\ $f_{\textrm{\tiny DM}} = f_{\textrm{\tiny DM}}(\mathcal{E}, L)$. The DM particles in such distributions follow orbits that can be either radially or tangentially biased, depending on the sign of $\beta_{\textrm{\tiny DM}}(r)$. Since the latter can not be inferred from observations, we have very limited information regarding this quantity. $N$-body simulations within the $\Lambda$CDM paradigm find only mild departure from the isotropic limit~\cite{bib2006NewA...11..333H, bib2008MNRAS.388..815W, bib2011MNRAS.415.3895L, bib2012ApJ...752..141L, bib2012JCAP...10..049S, Vera-Ciro-ml-2014ita}; there is also the \emph{cusp slope-central anisotropy} theorem, which states that at the center of a system $2 \beta_{\textrm{\tiny DM}} \geq - \ \mathrm{d} \ln \rho / \mathrm{d} \ln r$~\cite{An-ml-2005tm}. In the following, we are going to study radially biased orbits by adopting the so-called \emph{Osipkov-Meritt model}~\cite{Osipkov-ml-1979,Merritt-ml-1985}. This choice corresponds to a central part of the halo that is isotropic, while the DM particle orbits become increasingly radial in the outskirt:
\begin{align}
\label{eq:beta_osipkov_meritt}
\beta_{\textrm{\tiny DM}}(r) = \frac{r^2}{r^2 + r_a^2} \,,
\end{align}
where $r_a$ is the anisotropy scale radius. Even in this case the PSDF still depend on a single quantity, namely the variable $Q \equiv \mathcal{E} - L^2 / (2 r_a^2)$ and consequently eq.~\eqref{eq:density} can be inverted analogously to the Eddington's case~\cite{Osipkov-ml-1979,Merritt-ml-1985}:
\begin{align}
\label{eq:f_Osipkov-Merritt}
f_{\textrm{\tiny DM-OM}}(Q) =\; &  \frac{1}{\sqrt{8} \pi^2} \frac{d}{d Q} \int_{0}^{Q} \frac{d \Psi}{\sqrt{Q - \Psi}} \frac{d \rho_{\textrm{\tiny DM-Q}}}{d \Psi} \nonumber \\
=\; & \frac{1}{\sqrt{8} \pi^2} \left[ \int_{0}^{Q} \frac{d \Psi}{\sqrt{Q - \Psi}} \frac{d^2\rho{\textrm{\tiny DM-Q}}}{d \Psi^2} - \left. \left( \frac{1}{\sqrt{Q - \Psi}} \frac{d \rho_{\textrm{\tiny DM-Q}}}{d \Psi} \right) \right\rvert_{\Psi = 0} \right] \, ,
\end{align}
where one needs to rescale the density as $\rho_{\textrm{\tiny DM-Q}}(r) \equiv \rho_{\textrm{\tiny DM}}(r) \cdot (1 + r^2 / r_a^2)$. 

Finally, in this study we also consider the case of DM particles following circularly biased orbits. Here the central anisotropy theorem is automatically fulfilled for all considered density profiles and one can adopt the simplifying assumption of constant velocity anisotropy $\beta_\dm(r) = \beta_c$. The simplest choice for such a setup is the one, in which the PSDF is factorized as~\cite{bt08}:
\begin{align}
	f_{\textrm{\tiny DM}-\beta_c}(\mathcal{E}, L) = \left( \frac{L}{L_0} \right)^{-2 \beta_c} \cdot g_{\beta_c}(\mathcal{E}, L_0)
\end{align}
The procedure to invert eq.~\eqref{eq:density} is again similar to the one of Eddington's formula. It becomes particularly simple for $\beta_c = - 1/2$ when $g_{\beta_c}$ takes the following form:
\begin{align}
	\label{eq:psdf_beta_c}
	g_{-1/2}(\Psi, L_0) = \frac{L_0}{2 \pi^2} \frac{\mathrm{d}^2}{\mathrm{d} \Psi^2} \bigg(\frac{\rho_{\textrm{\tiny DM}}}{r} \bigg).
\end{align}
Analytic solutions can be also derived for other half-integer values of $\beta_c$, but the inversion could even be performed for any arbitrary constant value using Abel integral transform. We however focus only on the case of $\beta_c = - 1/2$, since it is sufficient for exploring the trend of Sommerfeld-enhanced annihilations in systems with circularly biased orbits. We indeed find significant $J$-factor boosts compared to the isotropic case already by considering such moderate values of orbital anisotropy.\footnote{We explicitly checked that by going to lower values of $\beta_c$ one finds even larger enhancement.}

\subsection{Comparing the velocity distributions}\label{sec:vel_dist}

\begin{figure}
	\centerline{
		\includegraphics[scale=.77]{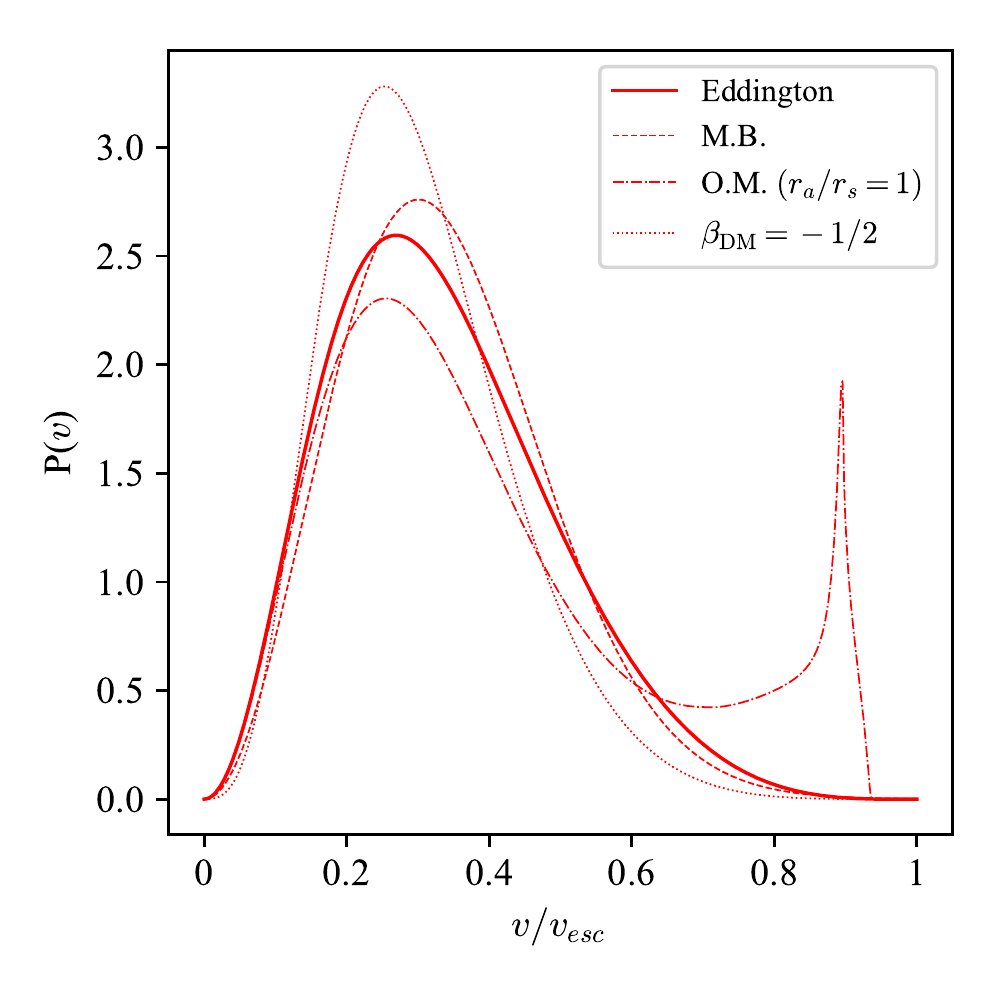}
		\includegraphics[scale=.77]{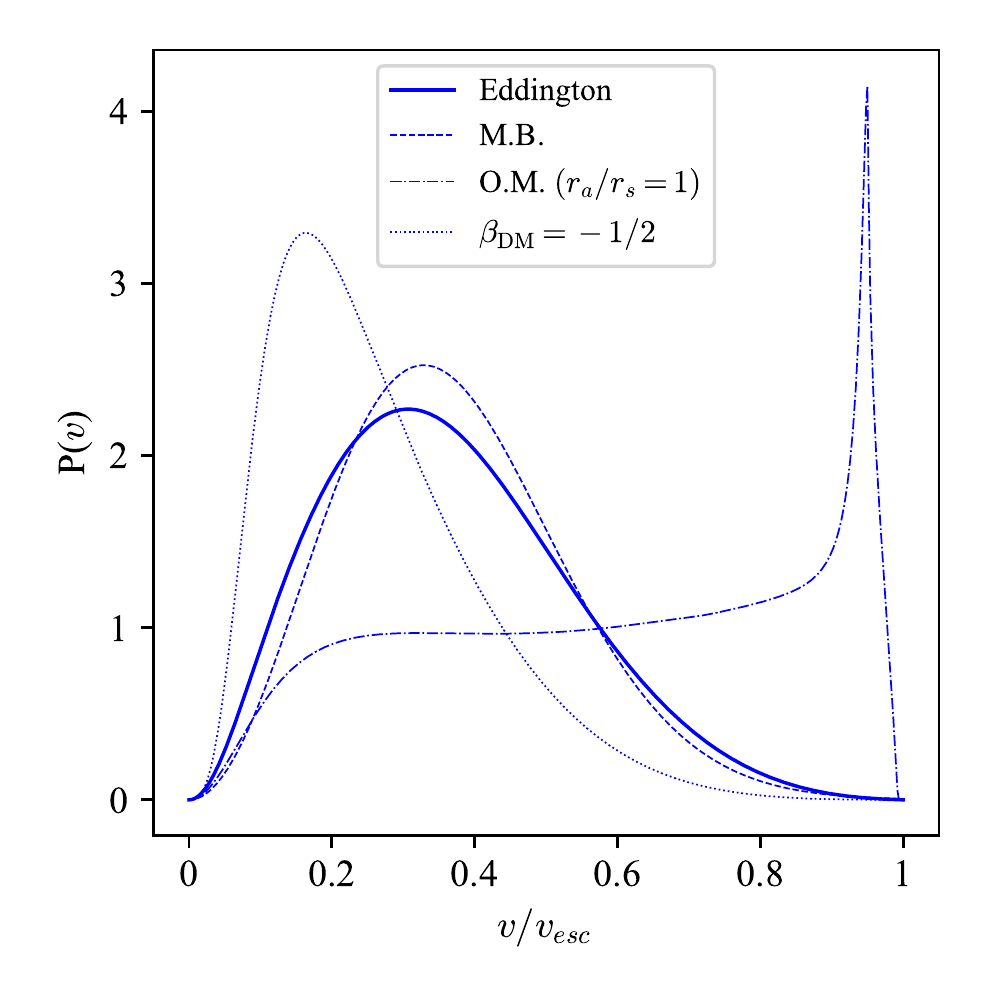}
	}
\vspace{-4mm}
	\caption{{Velocity probability distribution for various PSDF models. The left (right) plot corresponds to a NFW (Burkert) DM density profile; in both cases results are shown for $r / r_s = 0.3$, with $r_s$ being the scale radius. The velocities are normalized to the escape velocity $v_{\textrm{esc}}$.}}
	\label{fig:vel_dist}
\end{figure}

The four PSDF models listed above are all based on a given DM density profile and its corresponding gravitational potential, therefore one should obviously recover the same initial $\rho_\dm(r)$ after applying eq.~\eqref{eq:density}. However, the differences among them can be appreciated by looking at the corresponding velocity probability distributions:
\begin{align}
\mathrm{P}(v; r) = \frac{v^2}{\rho_{\textrm{\tiny DM}}(r)} \int d \Omega_{v}  f_{\textrm{\tiny DM}}(r, \vec{v}) \,,
\end{align}
where $d\Omega_v$ denotes the integral over the direction of the  velocity vector $\vec{v}$ (of course, such integral reduces to a  factor of $4 \pi$ for isotropic PSDFs).  In figure~\ref{fig:vel_dist} we show P$(v; r)$ for the four PSDFs under assumption of the two parametric density profiles --- a NFW on the left panel and a Burkert profile on the right --- at a fixed ratio of $r / r_s = 0.3$, chosen as representative radius since in both cases a significant contribution towards the total $J$-factor originates around this portion of the density profile, as discussed in section~\ref{sec:J_factors}. We can see that the standard isotropic modelling, using Eddington's inversion formula or Maxwell-Boltzmann approximation, yields similar P$(v; r)$, with the MB approach typically predicting slightly warmer particles, which however becomes increasingly significant with decreasing $r / r_s$. On the other hand, anisotropic PSDFs have distinct velocity distributions and exhibit consistent trends in comparison with the isotropic case for both $\rho_\dm(r)$ considered. Osipkov-Merritt model yields velocity distributions with more power at high velocities\footnote{The secondary peak close to $v_{\textrm{esc}}$ arises in  connection to the radial truncation of the profile. There are a few  possibilities on how to introduce it, see appendix~\ref{app:boundary}  for details. The sharp peak in the plots appears when a smoothing  function is introduced, while it would be less pronounced for a sharp  cut-off. Due to the nature of Sommerfeld enhancement this truncation  artifact at high velocities has however no sizable impact on our  results.}, which can be heuristically understood by the fact that particles  on radial orbits reach their terminal velocities at the center of halo. On the contrary, for $\beta_{\textrm{\tiny DM}}(r) = - 1/2$, where the orbits are circularly biased, one finds significantly colder central velocity distribution. This can be explained by noting that the circular velocity scales as $v_{\textrm{circ}} \propto r^{(3-\gamma)/2}$, given a density profile with the central slope $\gamma$ (i.e.\ $\rho_\dm(r) \propto r^{-\gamma}$ for $r \ll r_s$).

The trends sketched here for single particle velocity distributions  are to some extent representative also of the scalings with the  relative velocity in particle pairs, which is the relevant quantity when  addressing $J$-factors in presence of velocity dependent annihilation  cross-sections. In the following section we apply our analysis to  observational data of dSphs and examine the implications of various  phase-space distribution models.

\section{Probing DM phase-space distribution in the classical dSphs} \label{sec:bayesian}

Here we illustrate the main results of the paper. First, we report the details on the statistical analysis carried out on the available kinematic dataset of MW classicals in order to constrain the DM halo density profile in dSphs.
Note that the first part of the this section refers explicitly to parametric profiles.
In the second part we compute the $J$-factor for the MW classical dwarfs in the presence of velocity-dependent effects such as Sommerfeld enhancement.
The results for non-parametric profiles will be also given in the end for comparison.
We highlight the role of the DM velocity distribution assumption, quantifying the differences among the several cases considered, and show in particular the impact of anisotropic DM PSDFs against the more commonly adopted  isotropic phase-space distributions, according to the modeling  in eq.~\eqref{eq:f_Osipkov-Merritt}--\eqref{eq:psdf_beta_c} involving $\beta_{\textrm{\tiny DM}} \neq 0 $.

\subsection{Bayesian inference from dSph stellar kinematics}\label{sec:MCMC}

The small inferred ellipticities and the lack of evidence in favor of tidal disruption in the eight classical dSphs give reasonable support to our assumptions of spherical symmetry and dynamical equilibrium. Within such framework we can apply the standard spherical Jeans analysis, see e.g.~\cite{Charbonnier-ml-2011ft,Geringer-Sameth-ml-2014yza,Bonnivard-ml-2015xpq}.
We start by solving eq.~\eqref{eq:SphJeansEq} for the stellar component of these systems, for which $p_{r \, \star}(r)$ can be generally written as:
\begin{equation}
\label{eq:SolJeansEq}
p_{r \, \star}(r) = G_{N} \int_{r}^{\infty} d x  \, \frac{\rho_{\star}(x) M_{\textrm{tot}} (x)}{x^{2}} \exp\left[{2 \int_{r}^{x} dy \,  \frac{\beta_{\star}(y)}{y}}\right]  \,,
\end{equation}
where $G_{N}$ is the gravitational constant; the total mass profile of the system can be  approximated to the one of the DM component only, $M_{\textrm{tot}} \simeq M_{\textrm{\tiny DM}}$, given the large mass-to-light ratio exhibited by these objects. To make contact with data, eq.~\eqref{eq:SolJeansEq} is projected along the l.o.s.according to:
\begin{equation}
\label{eq:ProjJeansEq}
\sigma^2_{\rm los}(R) = \frac{1}{\Sigma_{\star}(R)} \int_{R^{2}}^{\infty} \frac{dr^{2}}{\sqrt{r^{2}-R^{2}}} \left( 1-\beta_{\star}(r) \frac{R^{2}}{r^{2}} \right) p_{r \, \star}(r)  \,,
\end{equation}
where we have introduced the surface brightness profile of the system, $\Sigma_{\star}(R)$. Note that the stellar density can be traced via an Abel transform of the surface brightness under the approximation of a constant luminosity profile for the stars in the galaxy. Once eq.~\eqref{eq:SolJeansEq} is plugged into eq.~\eqref{eq:ProjJeansEq}, such normalization drops out. For an adequate description of the surface brightness profile of the classical dwarfs, we rely on the Plummer model~\cite{bib1911MNRAS..71..460P,bib1987MNRAS.224...13D}, characterized by the projected half-light radius $R_{1/2}$, which provides good fits to the available photometric dataset for these objects~\cite{doi-ml-10.1093-ml-mnras-ml-277.4.1354}:
\begin{equation}
\Sigma_{\star}(R) \propto \left(1 + (R/R_{1/2})^{2} \right)^{-2} \ \ \Leftrightarrow \ \  \rho_{\star}(r) \propto 3/(4 R_{1/2}) \left(1 + (R/R_{1/2})^{2} \right)^{-5/2} \ .
\end{equation}
Notably, in the present study we include the two independent sources of uncertainty lying in the observational determination of $R_{1/2}$ (which is to a very good approximation given by $R_{1/2} \simeq \alpha_{1/2} D$), namely the error from determination of the heliocentric distance of the object, $D$, and the one from analysis of the photometric data when determining the angular half-light distance, $\alpha_{1/2}$. We adopt here for the eight classical MW satellites nominal values $\overline{D}$, $\overline{\alpha}_{1/2}$ and corresponding estimated errors $ \Delta{D}$, $\Delta{\alpha}_{1/2}$ reported in the compilation of ref.~\cite{bib2012AJ....144....4M}.

In order to predict l.o.s. velocity dispersion profiles, eq.~(\ref{eq:SolJeansEq})--(\ref{eq:ProjJeansEq}) require the modeling of the stellar orbital anisotropy and the DM halo mass of the galaxy. In an attempt to derive conservative bounds on DM halo parameters, we avoid the use of over-simplified stellar anisotropy profiles,  e.g.\ a spatially constant parameter~\cite{Strigari-ml-2006rd,Charbonnier-ml-2011ft,Geringer-Sameth-ml-2014yza}. In light of the poor indications concerning $\beta_{\star}(r)$ in dSphs, both from the side of N-body simulations~\cite{Campbell-ml-2016vkb,bib2017MNRAS.472.4786G} and present observations~\cite{bib2018NatAs...2..156M}, we rather advocate a 3-parameter fiducial model of the form:
\begin{equation}
\label{eq:beta_rad}
\beta_{\star}(r) = \frac{\beta_{0} + \beta_{\infty} (r/r_{\beta})^{2}}{1+(r/r_{\beta})^{2}} \,,
\end{equation}
i.e.\ the Baes-Van Hese proposal~\cite{Baes-ml-2007tx}, characterized by a transition from an inner regime governed by $\beta_{0}$ to an outer one set by $\beta_{\infty}$,  with characteristic scale $r_{\beta}$ and slope $\eta_{\beta} = 2$, which we for simplicity keep constant throughout our analysis.

The adopted combination of stellar Plummer model, stellar velocity dispersion anisotropy in eq.~(\ref{eq:beta_rad}), and the cuspy/cored DM halo profile defined in eq.~(\ref{eq:rho_NFW})--(\ref{eq:rho_BUR}), fully characterizes our study of dSph galactic dynamics with the spherical Jeans equation. The test-statistic we define in order to perform our analysis  on the measured stellar kinematics in MW satellites reads as follows:
\begin{equation}
\label{eq:loglikehood_kin}
 \mathcal{L}_{\textrm{kin}}  \equiv  \prod_{k = 1}^{\textrm{N}} \frac{1}{\sqrt{2 \pi } \, \Delta \sigma_{los \, (k)}\left(\alpha_{(k)}\right)} \exp \left[ -\frac{1}{2}   \left(  \frac{\overline{\sigma}_{los \, (k)}-\sigma_{\rm los}\left(\alpha_{(k)}\right)}{\Delta \sigma_{los \, (k)}\left(\alpha_{(k)}\right)}\right)^{2} \right]   \ .
\end{equation}
The above likelihood is suitable for a binned data analysis of dSph kinematics, see for instance~\cite{Bonnivard-ml-2015xpq}. For each bin $k \leq N$, with angular annulus $\alpha_{(k)} \simeq R_{(k)}/D$, we can compare theory predictions, $\sigma_{\rm los}\left(\alpha_{(k)}\right)$, against spectroscopic measurements, denoted here by $\overline{\sigma}_{los \, (k)}$; in doing so, we also take into account the observational uncertainty on the dataset binning, namely:
\begin{equation}
\label{eq:std_sigma_los}
 \Delta \sigma_{los \, (k)}\left(\alpha_{(k)}\right) \equiv \sqrt{ \left(\delta \sigma_{los \, (k)} \right)^{2} + \frac{1}{4}\left[\sigma_{\rm los}\left( \alpha_{(k)} + \Delta \alpha_{(k)}\right) - \sigma_{\rm los}\left(\alpha_{(k)} - \Delta \alpha_{(k)}\right) \right]^{2} } ,
\end{equation}
where $\delta \sigma_{los \, (k)} $ corresponds to the observational error  stemming from the spectroscopic measurement of the l.o.s. velocity dispersion, while $\Delta \alpha_{(k)}$ stands for the angular distance uncertainty associated with the k-$th$ bin.
Equipped with eq.~(\ref{eq:loglikehood_kin}), we proceed performing a Markov Chain Monte Carlo (MCMC) analysis exploiting the stellar kinematic dataset presented in~\cite{Geringer-Sameth-ml-2014yza}.\footnote{We are deeply grateful to M.G.~Walker, who has provided us stellar l.o.s. velocity dispersions for the classical MW satellites in bins of angular annuli. We wish to refer to~\cite{Walker2009,Geringer-Sameth-ml-2014yza} and more specifically to~\cite{Walker-ml-2008ax,bib2008ApJ...675..201M,bib2015MNRAS.448.2717W,bib2017ApJ...836..202S} for the details on the compilation of  the spectroscopic measurements characterizing the dataset analyzed in this work.}
Our fitting procedure is carried out along the lines of Bayes' theorem:
\begin{equation}
\mathcal{P}\left(\vec{\theta} \ | \ \textrm{data} \right) \propto \mathcal{P}_{0}\left(\vec{\theta} \, \right) \mathcal{L}_{\textrm{tot}}\left( \textrm{data} \ | \ \vec{\theta} \, \right) \,,
\end{equation}
where the posterior probability density function (p.d.f.) is sampled from the product of the prior probability distribution assigned to the set of model parameters $ \vec{\theta}$, with the  likelihood function reported in eq.~(\ref{eq:loglikehood_kin}), up to the overall normalization defining the so-called  evidence of the model (independent on $\vec{\theta} \ $). The general  model under scrutiny by means of Bayesian inference is defined by seven parameters:
\begin{equation}
\label{eq:model_params}
\vec{\theta} = \{ \rho_{s}, r_{s}, r_{\beta}, \beta_{0}, \beta_{\infty}, \alpha_{1/2}, D\} \ ;
\end{equation}
we explore the model parameter space restricting to the following set of ranges:
\begin{eqnarray}
\label{eq:MCMC_priors}
-5 \leq  \tilde{\rho}_{s} & \equiv &  \log_{10}  \left( \rho_{s}/ \textrm{[GeV cm$^{-3}$}\textrm{]} \right)  \leq 5 \,, \nonumber \\
-5 \leq  \tilde{r}_{s} & \equiv & \log_{10}  \left( r_{s}/ \textrm{[kpc]} \right)  \leq 2  \,, \nonumber \\
-3 \leq  \tilde{r}_{\beta} & \equiv &\log_{10} \left( r_\beta/ \textrm{[kpc]} \right)  \leq 1  \,,  \\
1 \leq   b_{0} & \equiv & 2^{\beta_{0}/(\beta_{0}-1)}   \leq 1.95 \,, \nonumber \\
0 \leq   b_{\infty} & \equiv & 2^{\beta_{\infty}/(\beta_{\infty}-1)}   \leq 1.95 \ . \nonumber
\end{eqnarray}
We assign flat prior distributions on the set $ \{ \tilde{\rho}_{s} , \tilde{r}_{s}, \tilde{r}_{\beta}, b_{0},  b_{\infty} \}$ according to the intervals reported in eq.~(\ref{eq:MCMC_priors}), while for the heliocentric distance $D$ and the half-light angle $\alpha_{1/2}$, we assume Gaussian prior with mean and standard deviation matching the corresponding observational information available, i.e.\ $\overline{D} \pm \Delta D$ and $ \overline{\alpha}_{1/2} \pm \Delta \alpha_{1/2}$ respectively.

\begin{figure}
\centerline{
\includegraphics[scale=0.33]{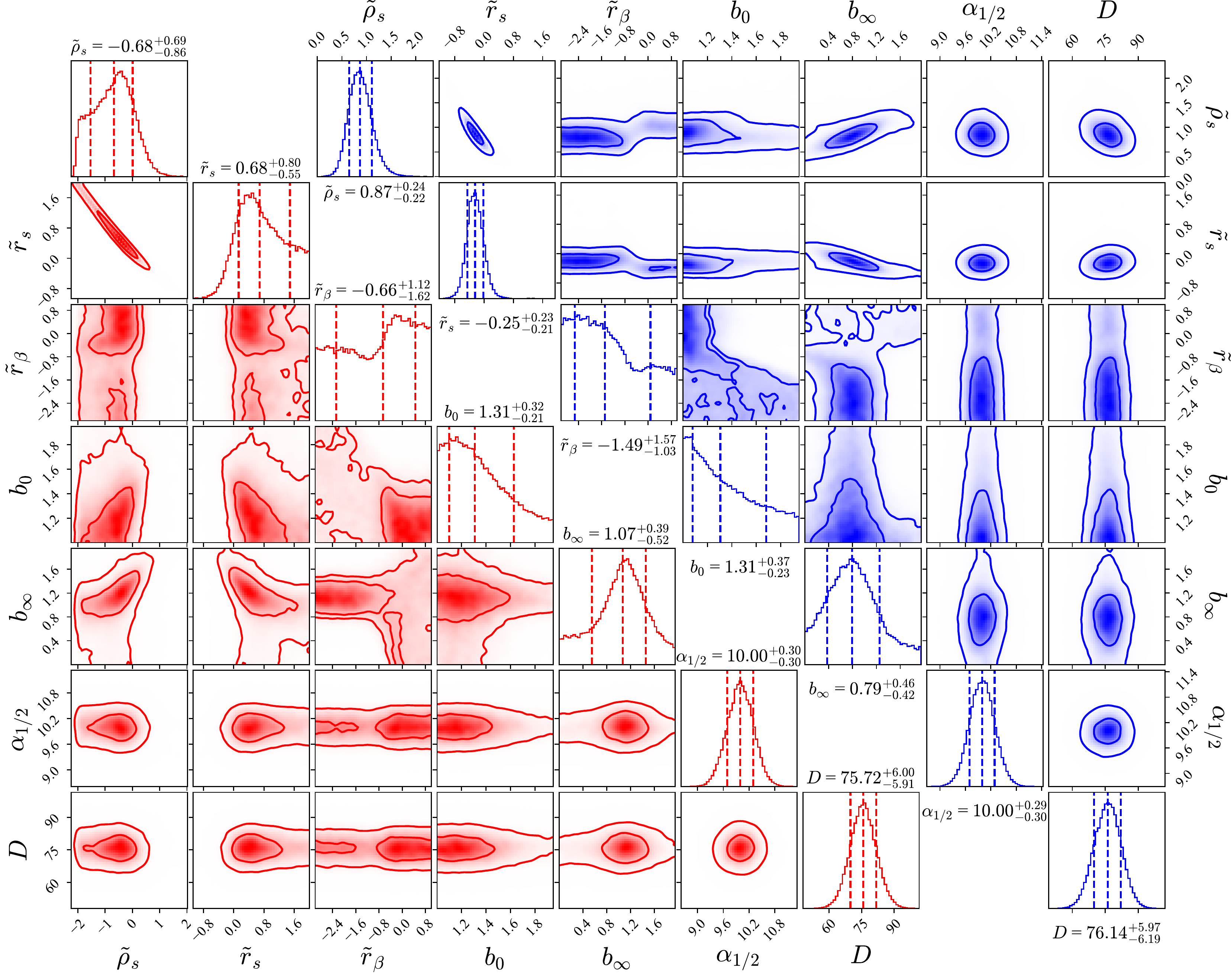}
}
\vspace{3mm}
\centerline{
\includegraphics[scale=0.33]{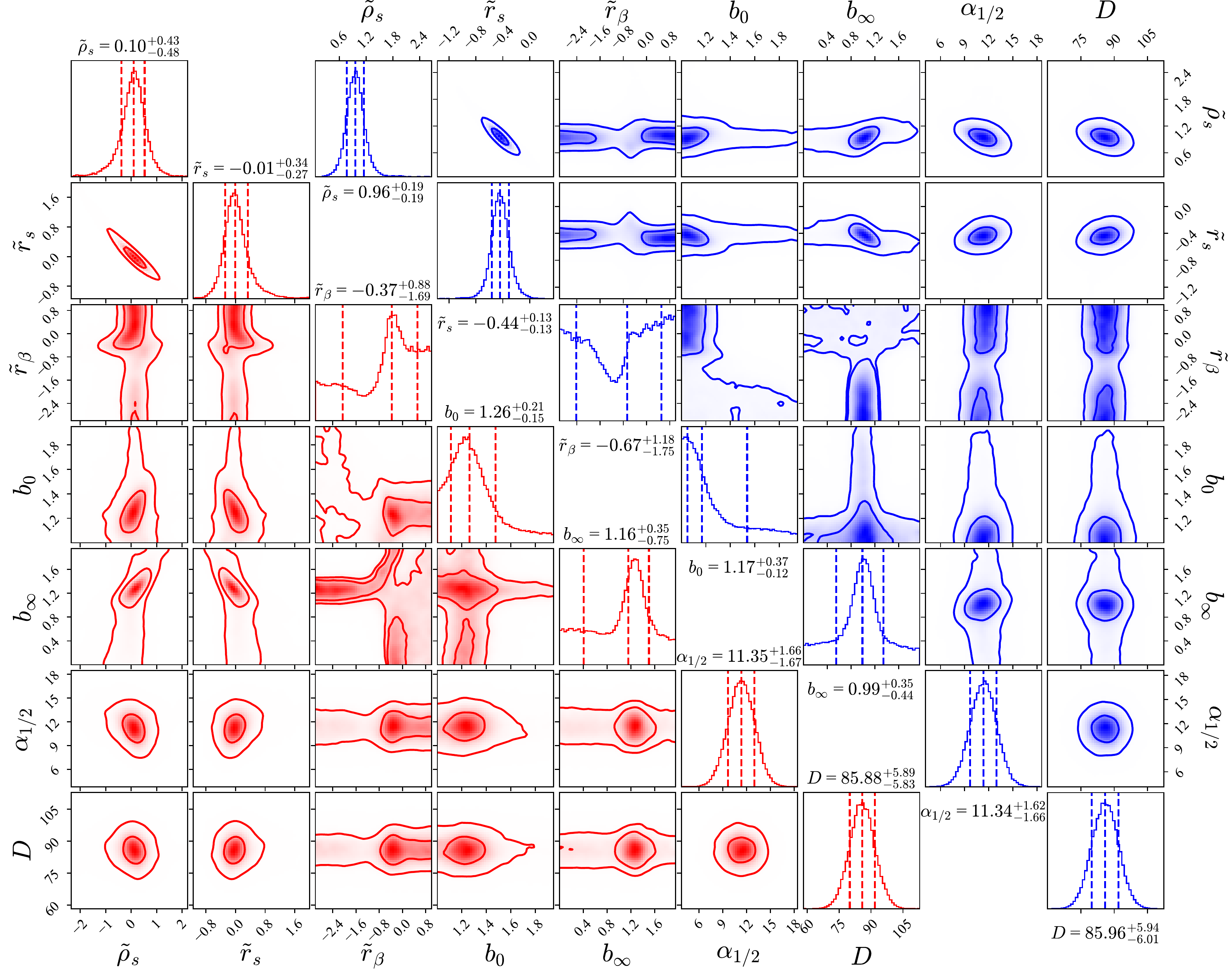}
}
\caption{{MCMC output of the estimated parameters from NFW and Burkert fits, respectively red and blue triangle plots, for Draco (upper panel) and Sculptor (lower panel). For each parameter, we report with dashed lines  the 16-$th$, 50-$th$, 84-$th$ percentile on the histogram of the marginalized posterior distribution. Correlations among the seven model parameters are also shown with the corresponding 68\% and 95\% highest probability regions. The parameter labels are defined as follows: $\tilde{\rho}_s = \log_{10} (\rho_s / \GeV \cm^{-3})$, $\tilde{r}_{x} = \log_{10} (r_{x} / \kpc)$ and $b_{x} = 2^{\beta_x / (\beta_x - 1)}$, while $\alpha_{1/2}$ and $D$ are in units of arcmin and kpc respectively.}}
\label{fig:corner_draco_sculpt}
\end{figure}

\begin{figure}
\centerline{
\includegraphics[scale=0.7]{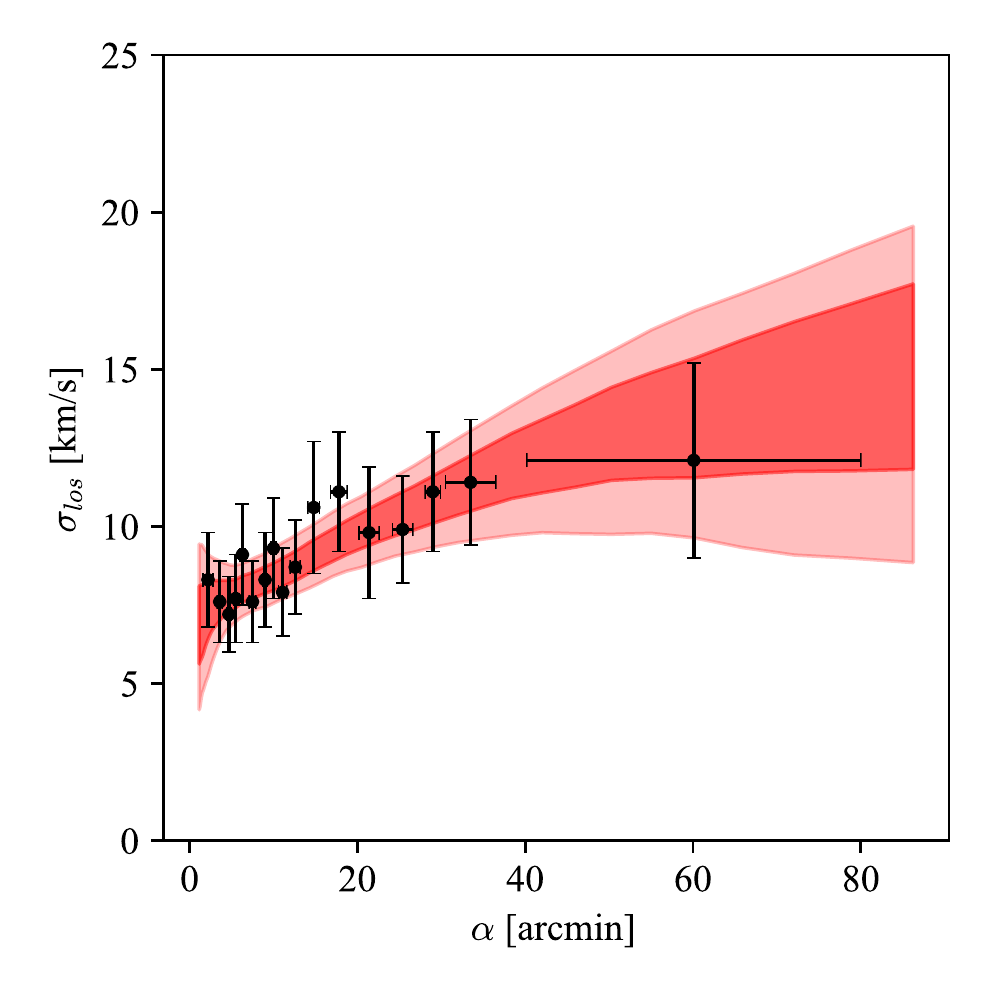}
\includegraphics[scale=0.7]{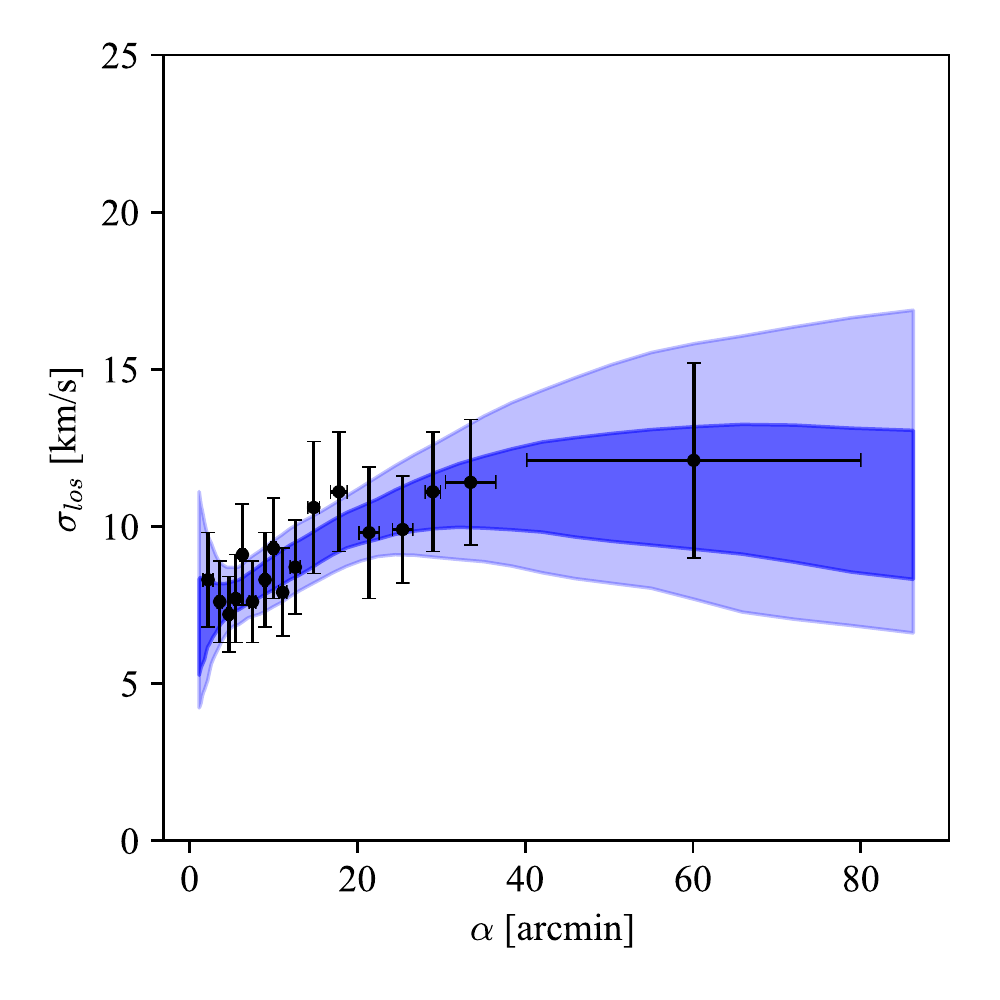}
}
\vspace{-2mm}
\centerline{
\includegraphics[scale=0.7]{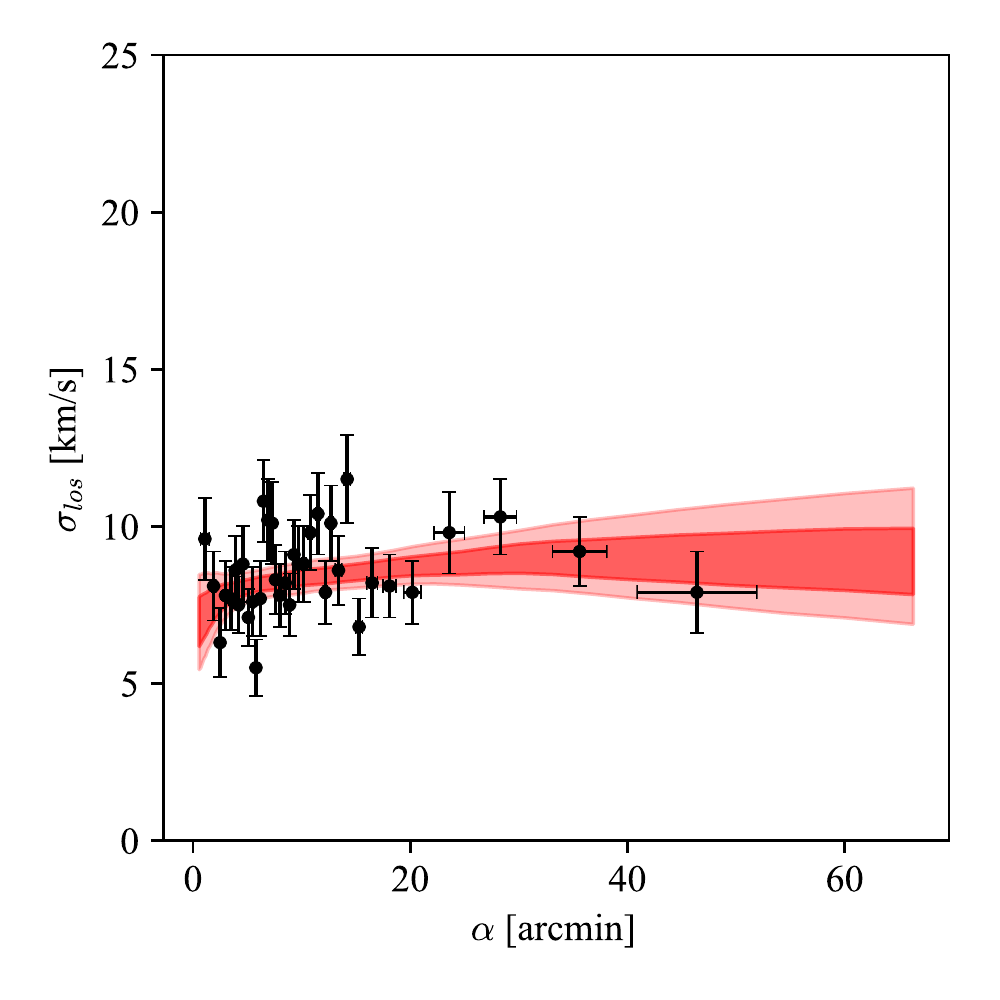}
\includegraphics[scale=0.7]{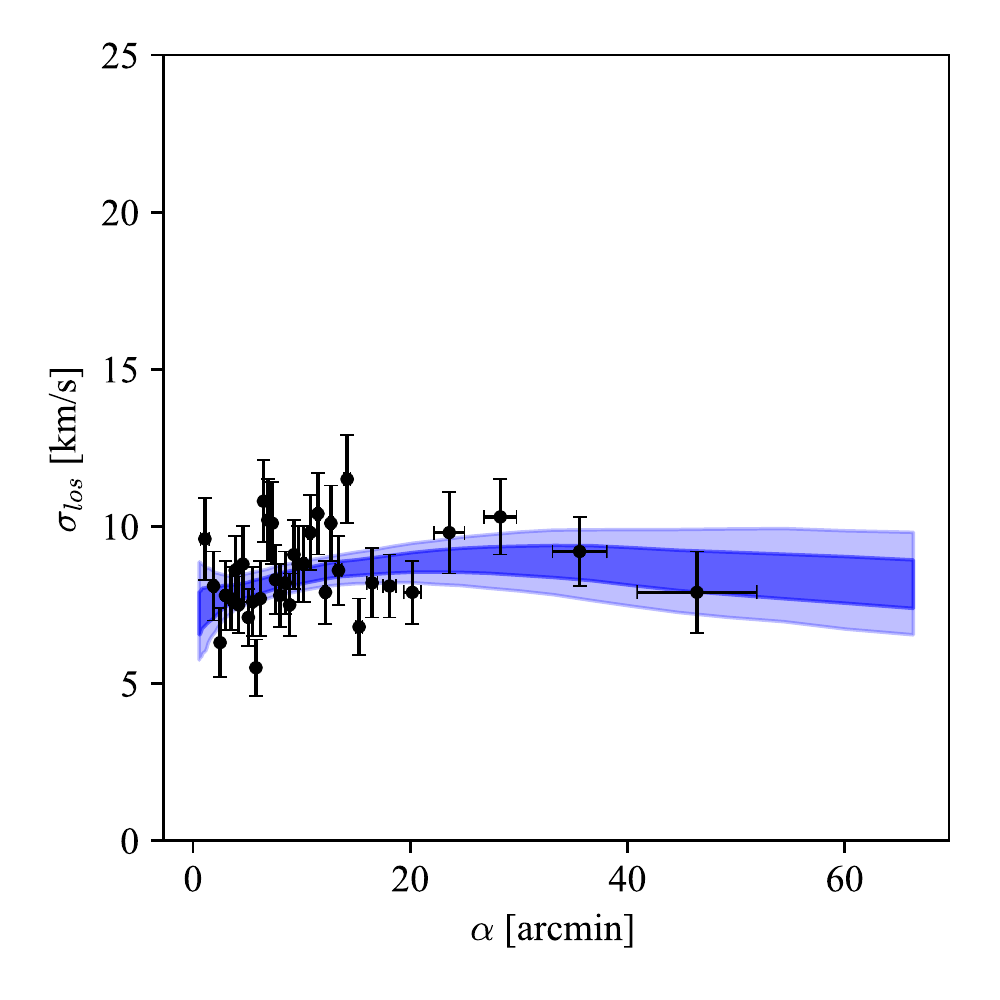}
}
\vspace{-3mm}
\caption{{Predicted l.o.s. velocity dispersion profile and related kinematic data for Draco (upper panels) and Sculptor (lower panels) in terms of the angular distance from the center of the dSph galaxy. In dark (light) color we show respectively the 68\% (95\%) highest density probability region obtained assuming a NFW (left panels) or a Burkert (right panels) halo profile when fitting the spectroscopic measurements.}}
\label{fig:slos_draco_sculpt}
\end{figure}

In light of the well-known mass-anisotropy degeneracy plaguing the spherical Jeans analysis~\cite{bib1982MNRAS.200..361B,Battaglia-ml-2013wqa,WalkerReviewMWdSphs,doi-ml-10.1093-ml-mnras-ml-stx1798}, some comments on the ranges appearing in eq.~(\ref{eq:MCMC_priors}) are in order. First, we wish to note that the adopted prior on $\tilde{r}_{s}$ involves a quite conservative upper-bound in relation to the expected size of a MW dSph halo --- reasonably well within $\mathcal{O}$(10$^{2}$) kpc --- while, for what concerns the lower-bound, smaller radii than that would probe extremely small scales, lower than $\mathcal{O}$(10$^{-2}$) pc. We remain quite agnostic also on the normalization of the DM halo, assigning a range to $\tilde{\rho}_{s}$ that covers ten dex in $\rho_{s}$. For what regards the orbital anisotropy parameters, we use the fact that by definition $\beta_{\star}(r) \leq 1$ and the result of the central-cusp anisotropy theorem~\cite{An-ml-2005tm},  restricting $\beta_{0,\infty}$ to vary in the allowed physical range. We further exploit the combination represented by $b_{0,\infty}$ introduced in eq.~(\ref{eq:MCMC_priors}) in order to equally weight tangential-like and radial-like stellar motion, however delimit the description of tangential orbits to values $\beta_{\star \,0,\infty} \gtrsim -25$ due to numerical limitations. Finally, we find it reasonable to restrict $r_{\beta}$ to the range essentially probed by the stellar kinematic dataset, namely $\mathcal{O}(1)$ pc--$\mathcal{O}(10)$ kpc.

To perform our MCMC analysis, we make usage of the \emph{emcee} package~\cite{ForemanMackey-ml-2012ig}, which implements the affine invariant algorithm of ref.~\cite{goodman2010} as the basic tool to build up the proposal distribution for the random walk of the chains. For each of the eight considered galaxies we let 500 walkers evolve for 2000 steps, starting from a neighborhood of the best-fit point in the seven dimensional parameter space, collecting a total of 10$^{6}$ samples. We remove the first half of them to account for the burn-in period and further check the auto-correlation length of the parameters in order to assess convergence in favor of independent draws from the posterior. As a final validation of our numerical analysis, we test the possible appearance of multi-modal solutions, which are difficult to sample within an ordinary MCMC sampling algorithm. In order to do that, we repeat the full analysis of the eight objects by applying the default nested sampling method available with the \emph{pymultinest} library~\cite{Buchner-ml-2014nha}, which implements the importance sampling algorithm proposed in ref.~\cite{Feroz-ml-2013hea}. We used 1000 live points subjected to the same prior distributions as discussed above and adopted the default tolerance of 0.5 for the estimated remaining evidence as a stopping criterion. For all the NFW and Burkert fits we found a remarkable agreement between the affine-invariant ensemble MCMC and the important nested sampling analyses, resulting in nearly identical posterior distributions.

In figure~\ref{fig:corner_draco_sculpt} we show the outcome of our Bayesian fit on Sculptor and Draco datasets, describing the underlying DM halo with the NFW (red triangle plot) or with the Burkert (blue triangle plot) profile. We report the one-dimensional marginalized posterior p.d.f. for each of the estimated parameters, highlighting the 16-$th$, 50-$th$, 84-$th$ percentiles, and their joint probability distribution with the 68\% and 95\% highest probability density (h.p.d.) contours. While a strong correlation between the DM halo parameters emerges in all our fits, we can observe for the case of Draco and Carina that the NFW scenario is sensitive to the physically motivated upper-limit assigned on $r_{s}$. As a general trend, we find in NFW fits a non-negligible correlation among DM halo parameters and $\beta_{0,\infty}$, while for the fits with Burkert profile the correlation of the halo parameters with inner trend of the stellar anisotropy gets essentially milder. Heliocentric distance and half-light angle show overall mild correlations in the bi-dimensional joint distributions with the rest of the fitted parameters. Finally, to provide a concrete picture of the goodness of our fits, we show in figure~\ref{fig:slos_draco_sculpt} the 68\% and 95\% h.p.d. interval for the predicted l.o.s. velocity dispersion profile in Draco and Sculptor, together with their binned  data points. As has been noted by previous authors~\cite{Walker-ml-2011fs, Charbonnier-ml-2011ft, Bonnivard-ml-2015xpq}, both cuspy and cored profiles can provide an optimal description of the present stellar kinematics in MW classical dwarfs, if one allows for a flexible enough stellar anisotropy profile.

\subsection[\texorpdfstring{$J$}{J}-factor estimates from PSDFs]{\texorpdfstring{\boldmath $J$}{J}-factor estimates from PSDFs}\label{sec:J_factors}

The main motivation of our paper is to accurately predict the expected prompt gamma-ray emission from DM annihilations in a general setting, where the relevant cross-section can be velocity dependent. As already motivated in the introduction, the net effect can be nicely encaptured in a velocity-averaged enhancement factor, namely:
\begin{equation}
\label{eq:J-integrand}
\langle S (v_{\textrm{rel}}) \rangle(r) = \frac{1}{\rho_\dm^2(r)} \int d \vec{v}_1 f_\dm(r, \vec{v}_1) \int d \vec{v}_2 \; f_\dm(r, \vec{v}_2)
S(|\vec{v}_{\textrm{rel}}|)\,.
\end{equation}
which reduces to 1 in case of velocity-independent annihilations. The reference case we have in mind is connected to the Sommerfeld effect~\cite{Sommerfeld-ml-1931}, which has been addressed by numerous studies in context of DM phenomenology, see e.g.~\cite{Boddy-ml-2017vpe,Lu-ml-2017jrh,Bergstrom-ml-2017ptx}. In this perspective, the key feature is the strong enhancement of the annihilation cross-section for highly non-relativistic (slow) particles that are charged under a force with light or massless mediator. The enhancement effect is quantum in its nature and stems from the distortion of the wave-function of the incoming particle states due to the exchange of a mediator sufficiently light to establish a regime of long-range interactions~\cite{Hisano-ml-2003ec,Hisano-ml-2004ds}. Sommerfeld-enhanced cross-sections for DM particle $\chi$ annihilating through the mediator $\phi$, with $m_{\chi} \gg m_{\phi}$, can be computed by solving a non-relativistic Schr\"{o}dinger equation with a potential related to long-range forces or by resummation of ladder $\phi$-exchanges in the diagrammatic field theory approach~\cite{Iengo-ml-2009ni}. Results are usually encapsulated as the ratio of wave functions in presence and absence of the long-raged force, i.e.\ $S = \left| {\chi(0)}/{\chi(\infty)}\right|^2$. While in general there is no analytical solution for the Yukawa potential, it can be very well approximated by Hulten's potential~\cite{Feng-ml-2009hw, bib2010JCAP...02..028S}, for which one finds:
\begin{equation}
\label{eq:sommerfeld}
S(v_{\textrm{rel}}; \, \xi) = \frac{\pi \alpha_{\chi}}{v_{\textrm{rel}}} \frac{\sinh \left( \frac{12 v_{\textrm{rel}}}{\pi \alpha_{\chi} \xi} \right)}{\cosh \left( \frac{12 v_{\textrm{rel}}}{\pi \alpha_{\chi} \xi} \right) - \cos \left( 2 \pi \sqrt{ \frac{6}{\pi^2 \xi} - \left( \frac{6 v_{\textrm{rel}}}{\pi^2 \alpha_{\chi} \xi} \right)^2} \right)} \,,
\end{equation}
where $\xi \equiv m_{\phi} / \left( \alpha_{\chi} m_{\chi} \right)$, with the long-range force strength dictated by the coupling constant $\alpha_{\chi}$. The above expression provides an enhancement of the DM annihilation rate that becomes negligible in the limit of large $v_{\textrm{rel}}$, or when $m_{\phi} \gtrsim m_{\chi}$, which implies $S(v_{\textrm{rel}}; \, \xi \gg 1) \rightarrow 1$. For small $v_{\textrm{rel}}$ there are two additional limiting behaviors. In case of vanishing mediator mass, which is often referred to as the Coulomb regime, one finds:
\begin{align}
\label{eq:Sommnonres}
S(v_{\textrm{rel}}; \, \xi \ll 1) \approx \frac{\pi \alpha_{\chi}}{v_{\textrm{rel}}}.
\end{align}
The third limiting behavior occurs at resonant values of $\xi_{\textrm{res}}$, where the enhancement becomes even stronger for low relative velocities. Form eq.~\eqref{eq:sommerfeld} one finds the following:
\begin{align}
\label{eq:Sommres}
S\bigg(v_{\rm rel} \; ; \; \xi_{\textrm{res}} = \frac{6}{\pi^2 n^2}\bigg) \approx \frac{\alpha_{\chi}^2}{v_{\textrm{rel}}^2 n^2} \;\;\; \textrm{for each} \;\;\; n \in \mathbb{N}.
\end{align}
In correspondence to these three regimes we will use a subscript notation for the $J$-factors, where $J_{\alpha-\textrm{\tiny X}}$ denotes its value in the $S \propto v_{\textrm{rel}}^{-\alpha}$ regime for phase-space model abbreviated by X (E for Eddington's inversion, MB for Maxwell-Boltzmann approximation, OM for Osipkov-Merritt model and $\beta_c$ for the $\beta_\dm(r) = -1/2$ case). At this point we also specify that all of our results were computed for $\alpha_{\chi} = 1 / 100$ and aperture of $\alpha = 0.5^{\circ}$ in the instrument acceptance cone $\Delta \Omega$, unless stated otherwise.

\begin{figure}
\centerline{
\includegraphics[scale=.77]{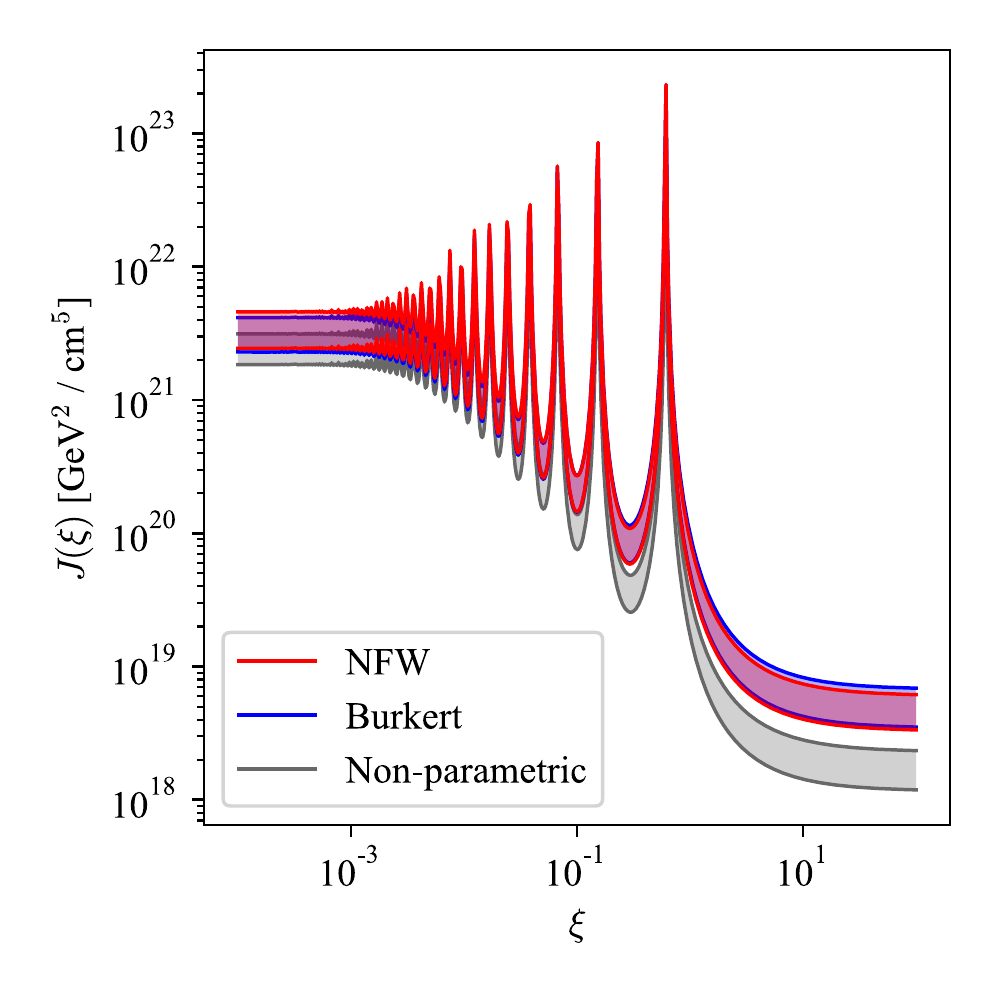}
\includegraphics[scale=.77]{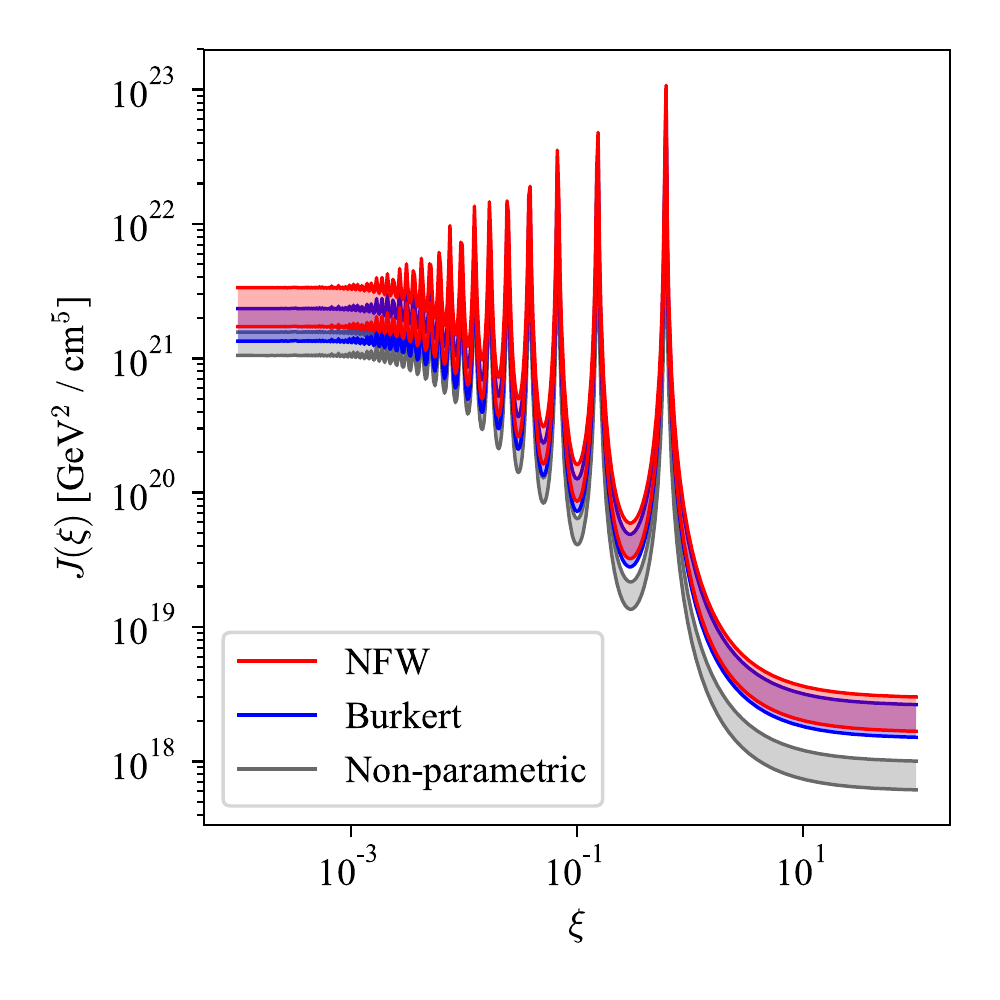}
}
\vspace{-4mm}
\caption{{$J$-factor dependence on the parameter $\xi \equiv m_{\phi} / \left( \alpha_{\chi} m_{\chi} \right)$; the bands displayed correspond to the 68\% highest density probability region obtained assuming a NFW, a Burkert or a non-parametric halo profile in case of Draco (left panel) and Sculptor (right panel).}}
\label{fig:J_xi}
\end{figure}

Figure~\ref{fig:J_xi} shows the dependence of $J$-factors on the combination of particle physics parameters $\xi$. The two panels refer again to the case of Draco (left panel) and Sculptor (right panel), while the three bands displayed are all computed under the assumption of isotropic velocities for PSDF given by Eddington's inversion formula: they correspond to the 68\% h.p.d. interval, derived from the above statistical analysis for the NFW and Burkert parametric profiles and by consistently propagating the errors on $R_{1/2}$, $D$ and averaged $\sigma_{\rm los}$ for the non-parametric profile, which was obtained via Jean's inversion procedure under assumption of flat stellar l.o.s. velocity dispersion and in the limit of circular stellar orbits, $\beta_{\star} \rightarrow - \infty$. Each of the three bands in the plots shows the three limiting regimes of the Sommerfeld effect: for large values of $\xi$ the standard, non-enhanced, values of $J$-factors are recovered. By decreasing $\xi$ one first encounters the resonance peaks, at which extremely large boost can be obtained, up to factors of $\mathcal{O}(10^5)$, with the peak at the largest $\xi$ (the one corresponding to $n=1$ in eq.~(\ref{eq:Sommres})) providing the largest enhancement.
By going to even lower values of $\xi$ one enters the Coulomb limit, where the enhancement saturates at factors of $\mathcal{O}(10^3)$; while the corresponding boost is notably smaller then on the resonances, this regime requires less fine-tuning on particle physics parameters.
In same plot the three bands clearly exhibit slight differences that arise among the considered density profiles. For both, Draco and Sculptor, we see larger net enhancement for cuspy density profiles (i.e. the NFW and our sample non-parametric case), since they typically imply deeper potential wells and therefore these halos host colder particle populations at their centers. At the same time, the effect one finds for a given dwarf cannot be rigidly applied to another object, since details of the enhancement depend on the preferred region in the parameter space. In general the larger the halo concentration, the larger the flux increase: e.g., in Draco we found that the fit in case of the NFW profile points to significantly larger $r_s$ and lower $\rho_s$ than for Sculptor, see figure~\ref{fig:corner_draco_sculpt}, while for the Burkert profile the preferred regions in parameter space are closer one to the other; correspondingly we find a smaller relative boost in the NFW versus Burkert comparison for Draco, while it is appreciably larger for Sculptor.

\begin{figure}
\centerline{
\includegraphics[scale=.72]{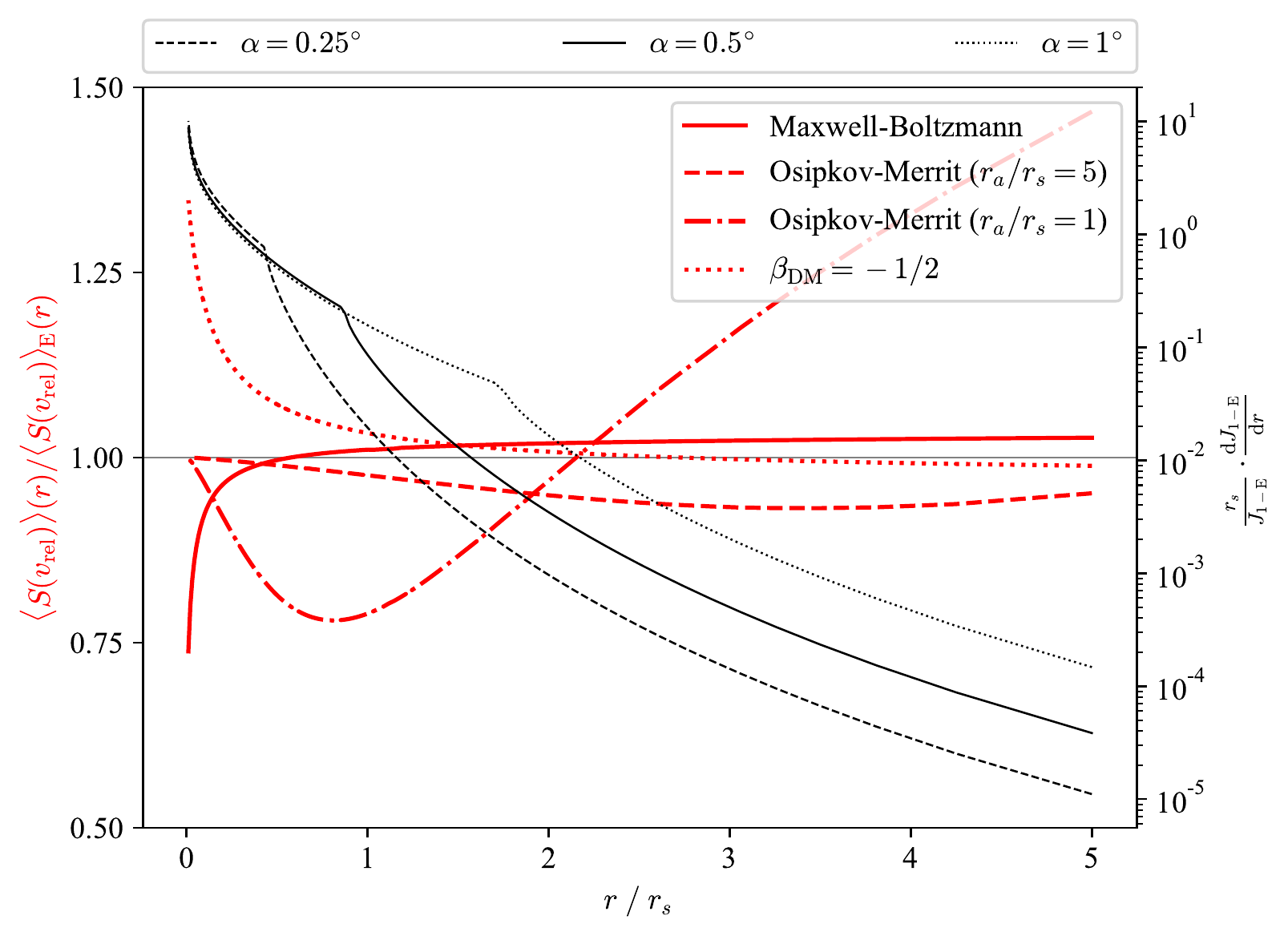}
}
\vspace{3mm}
\centerline{
\includegraphics[scale=.72]{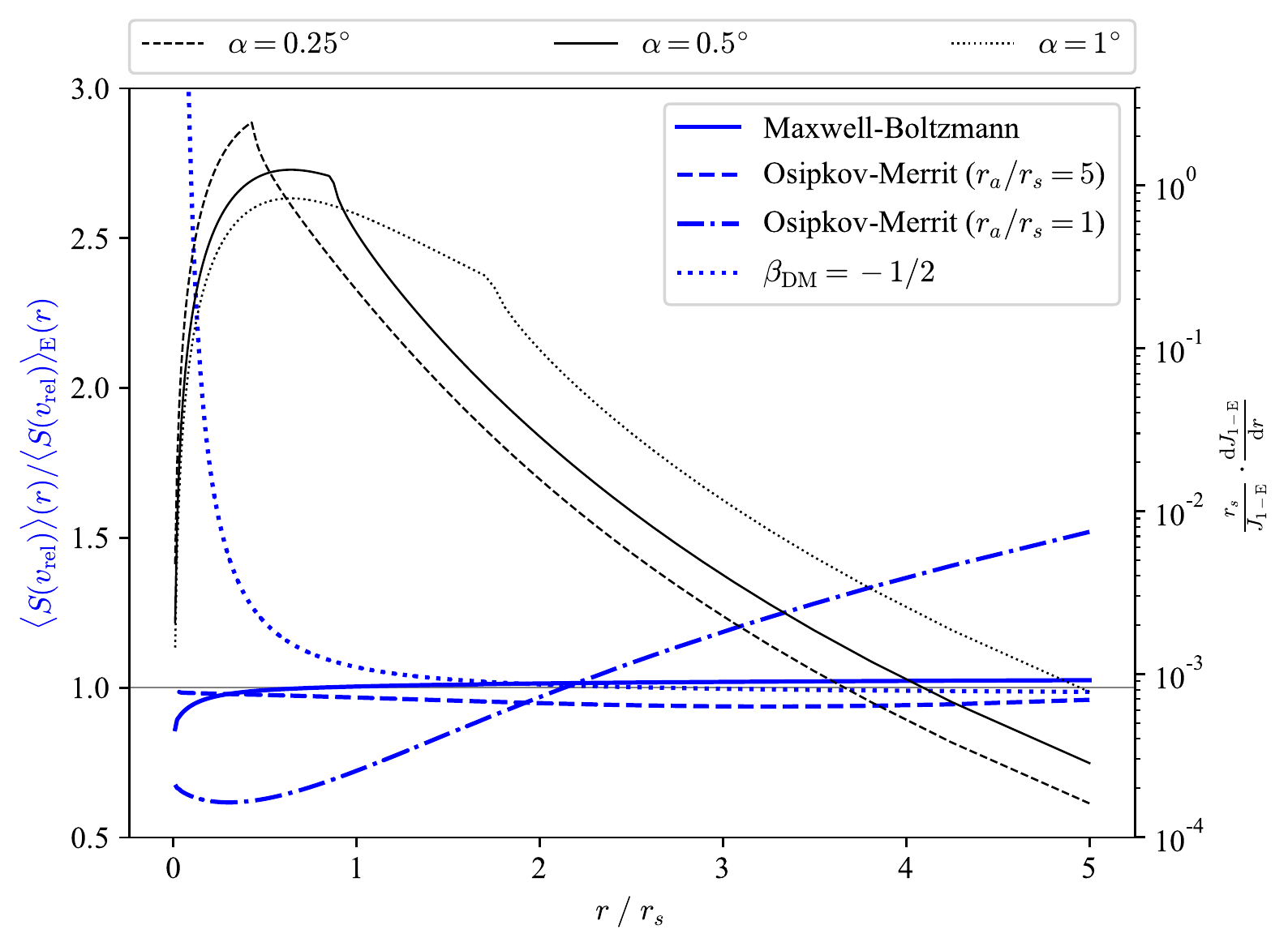}
}
\caption{{Rescaling of the velocity dependent factor $\langle S (v_{\mathrm{rel}}) \rangle$ in case of a given PSDF with respect to the result obtained assuming Eddington's inversion formula, as a function of the radial distance $r$ from the center of the halo (thick coloured curves). To illustrate where such rescalings are relevant, we also show the normalized integrands in the formula for the $J$-factor, assuming the Eddington's model and the Coulomb regime for the Sommerfeld-enhancement, when the $J$-factor for an instrument pointing towards the center of a spherical object is rewritten in terms of an integral over $r$; three different instrument angular apertures $\alpha$ are considered, having fixed the ratio of the object distance $D$  to the scale length $r_s$ to 100 (thin black lines). The upper (lower) panel assumes a NFW (Burkert) density profile for the DM halo.}}
\label{fig:J_integrand}
\end{figure}

As we wish to illustrate now, the choice of phase-space modelling has also a significant impact on the extrapolated enhancements, both when considering an approximation to the isotropic case, such as for the MB model introduced above, and even more drastically when one allows for anisotropic velocity distributions. In figure~\ref{fig:J_integrand} we consider the ratio $\langle S (v_{\mathrm{rel}}) \rangle (r) / \langle S (v_{\mathrm{rel}}) \rangle_{\textrm{\tiny E}} (r)$, comparing the velocity averaged enhancement of a given PSDF to the Eddington's case at the radial distance $r$ from the center of halo.
The angular and line-of-sight integrals appearing in the definition of $J$-factor, see eq.~(\ref{eq:j_factor}), when pointing towards the center of a spherical object, can be reduced to an integral on $r$, while performing the other integrals analytically, see~\cite{Ullio-ml-2016kvy}. After rewriting $J$ as $J = \int_0^{r_b}  dr \,dJ/dr(r)$, where $r_b$ is the outer truncation radius, the appropriately normalized ratio $dJ/dr$ over $J$ --- assuming the Eddington's PSDF and the Coulomb regime for the Sommerfeld-enhancement --- is plotted in figure~\ref{fig:J_integrand} with thin black lines (with corresponding vertical axis scale at the r.h.s.\ of the plots; note that this scale is logarithmic). The three lines in each plot are for three different apertures $\alpha$ of the instrument acceptance cone $\Delta\Omega$. The curves for the NFW profile peak all at $r=0$, while for the Burkert profile the largest contribution to $J$ shifts towards $r \approx r_s$; analogous trends --- but sharper --- would appear also in the resonant regime of the Sommerfeld enhancement. This check provides a visual guidance to the plots, indicating the intervals in $r$ over which there is a significant impact on the expected annihilation flux, would a different DM velocity distribution provide a significantly different $\langle S (v_{\mathrm{rel}}) \rangle (r)$. In figure~\ref{fig:J_integrand} thick lines show this quantity as computed for the MB, OM, $\beta_c$ PSDF over the result for the Eddington's case (now the reference vertical scale is on the l.h.s.\ of the plots). From the plot one can see that, for the NFW profile, the Maxwell-Boltzmann approximation gives a systematic (but numerical-friendly) underestimate of the true result in the isotropic case, since the corresponding curve is smaller than 1 at the peak of the contribution to $J_1$ (and even more so for $J_2$), while it gets marginally above 1 in the radial range less relevant for the $J$-factor; the MB approximation causes instead a much smaller error for the Burkert profile. In the same way one sees that even the very small bias on circular velocity introduced by the model with $\beta_\dm(r) = -1/2$, impacting on the abundance of slow particles at the center of the systems, is going to significantly boost  $J_{1,2-\beta_c}$ for both density profiles. Finally, for what regards Osipkov-Merritt model, $J_{1,2}$ tend to be smaller then in the isotropic case, which however becomes significant only when the suppression in $\langle S (v_{\mathrm{rel}}) \rangle (r)$, which is maximized at $r$ slightly below $r_a$, gets within the radial range relevant for the computation of $J_{1,2}$, namely if we consider $r_a$ close to $r_s$ for Burkert and $r_a \lesssim r_s$ for NFW profile.

\begin{figure}
	\centerline{
\includegraphics[scale=0.75]{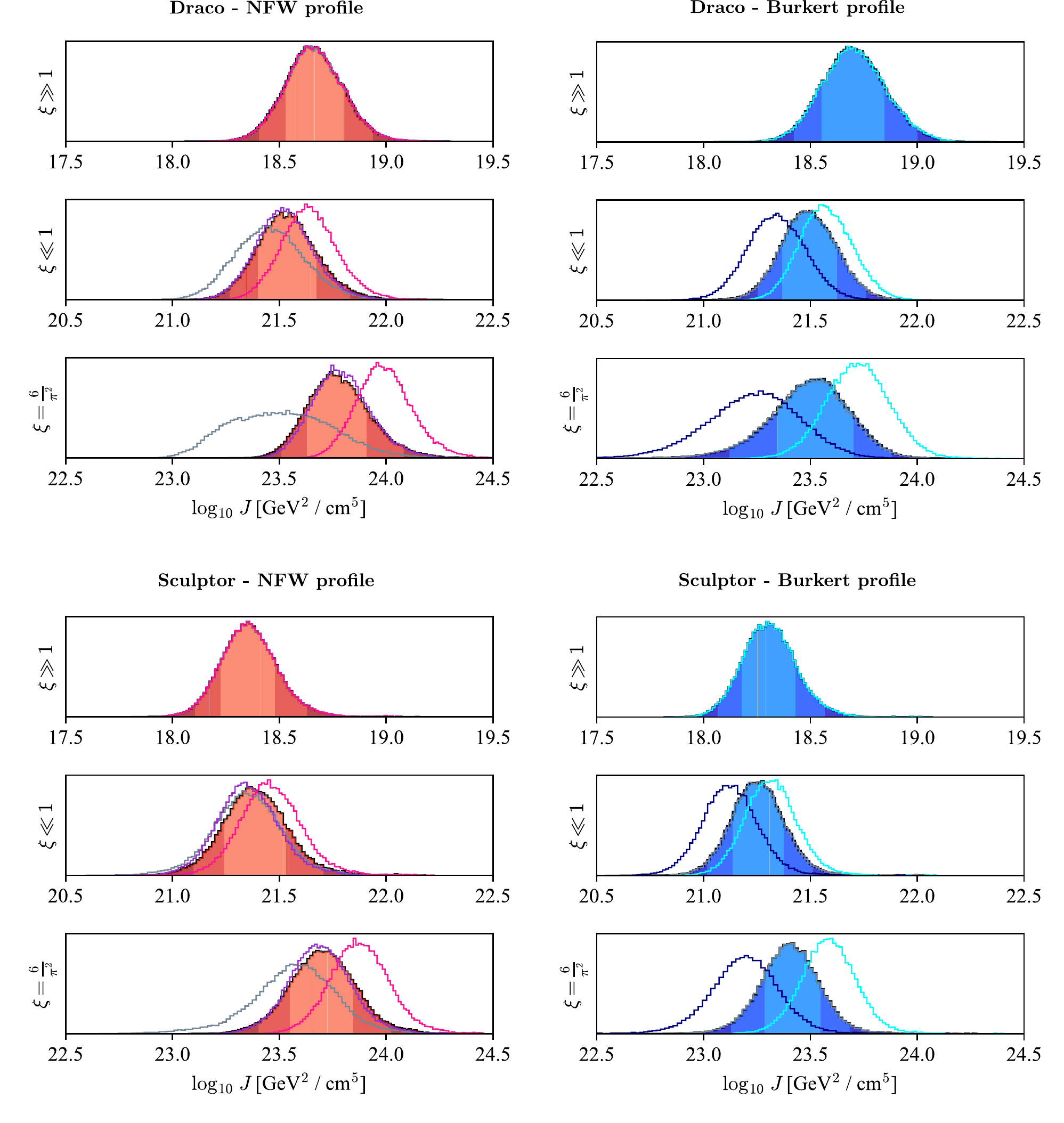}
}
\vspace{-3mm}
\caption{{Posterior distributions of $J$-factors in the three regimes of enhancement. The upper and lower figures are respectively representing the study cases of Draco  and Sculptor under the assumption of NFW (left) and Burket (right) DM density profiles. The histograms with colored 68\% and 95\% highest density probability regions were obtained using Eddington's inversion. Also, we report with gray lines the posterior related to the Maxwell-Boltzmann scenario, while with dark (purple for NFW and dark blue for Burkert) colored lines the Osipkov-Merritt model, while we show with light colored lines (pink for NFW and cyan for Burkert) the case of $\beta_{\textrm{\tiny DM}} = - \frac{1}{2}$ modelling.
}}
\label{fig:J_histogram}
\end{figure}

With the general trends delineated in figure~\ref{fig:J_integrand} , we can now quantify the Sommerfeld effect for Milky Way classical dSphs. From the samples obtained in the MCMC analysis described in section~\ref{sec:MCMC}, we can actually compute the $J$-factor posterior distributions. Note that evaluation requires the non-trivial computational task of performing multidimensional integration for each of the MCMC events collected. In order to make the demanding numerics feasible, we can resort to the scaling relations explained in greater detail in appendix~\ref{app:J-factor_scaling}. In figure~\ref{fig:J_histogram} we show the $J$-factor p.d.f. for our benchmark galaxies, Draco and Sculptor. The foreseen trend from figure~\ref{fig:J_integrand} is nicely met in our findings for $J$-factors, applied to dSph kinematic data. We see significant shifts in the peak of the distributions for the $\beta_c$ model, and for the MB model with NFW profile, as well as the for OM model in the Burkert case. We also note that the width of distributions depends on the PSDF under consideration, most clearly visible by the increased spread for MB approximation in NFW case and for OM model in the Burkert case. On the contrary, $\beta_c$ model yields slightly narrower distributions as the enhancement effect is simply dominated by the center of halo resulting in lesser dependence on the structural parameters.

\begin{figure}
\centerline{
\includegraphics[scale=0.78]{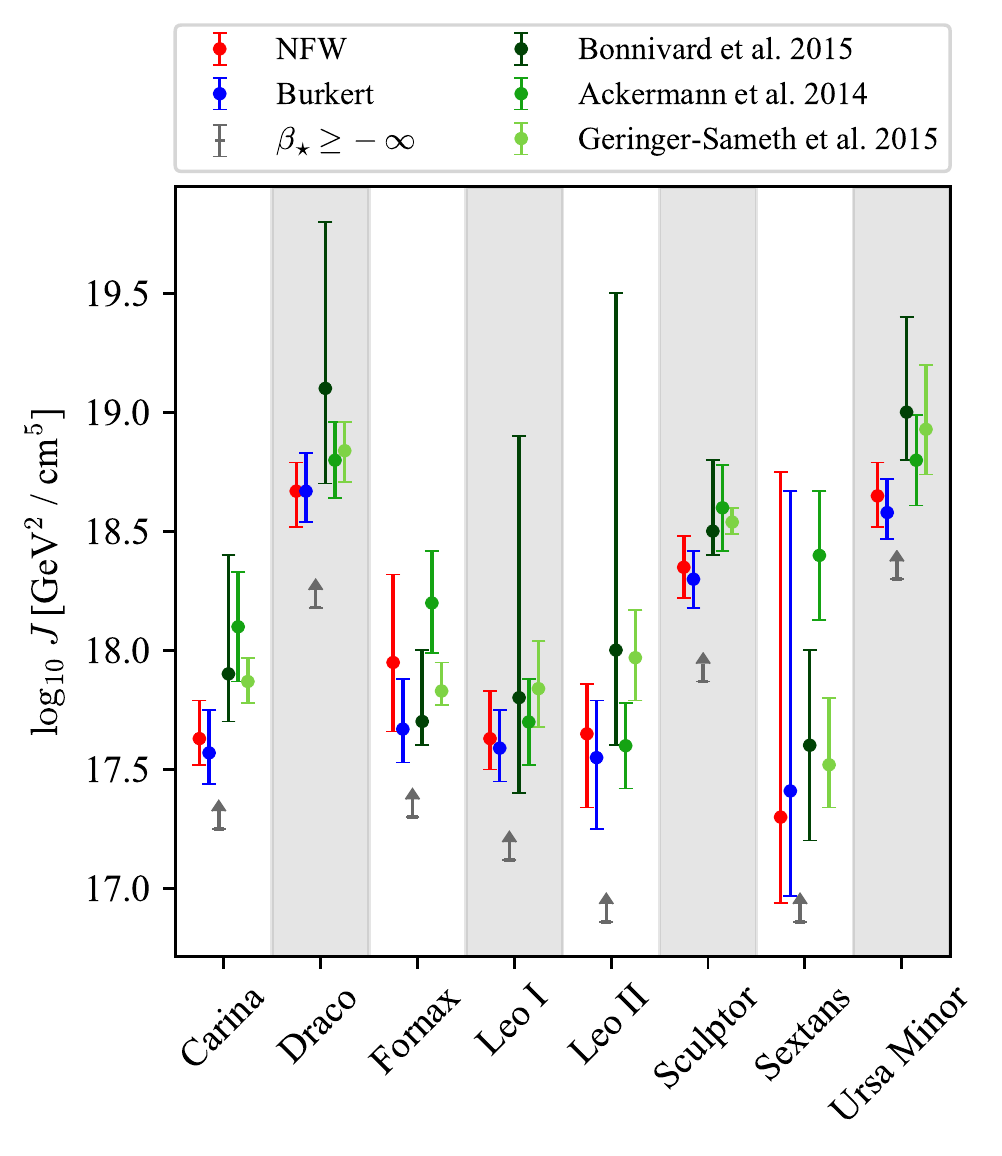}
\includegraphics[scale=0.78]{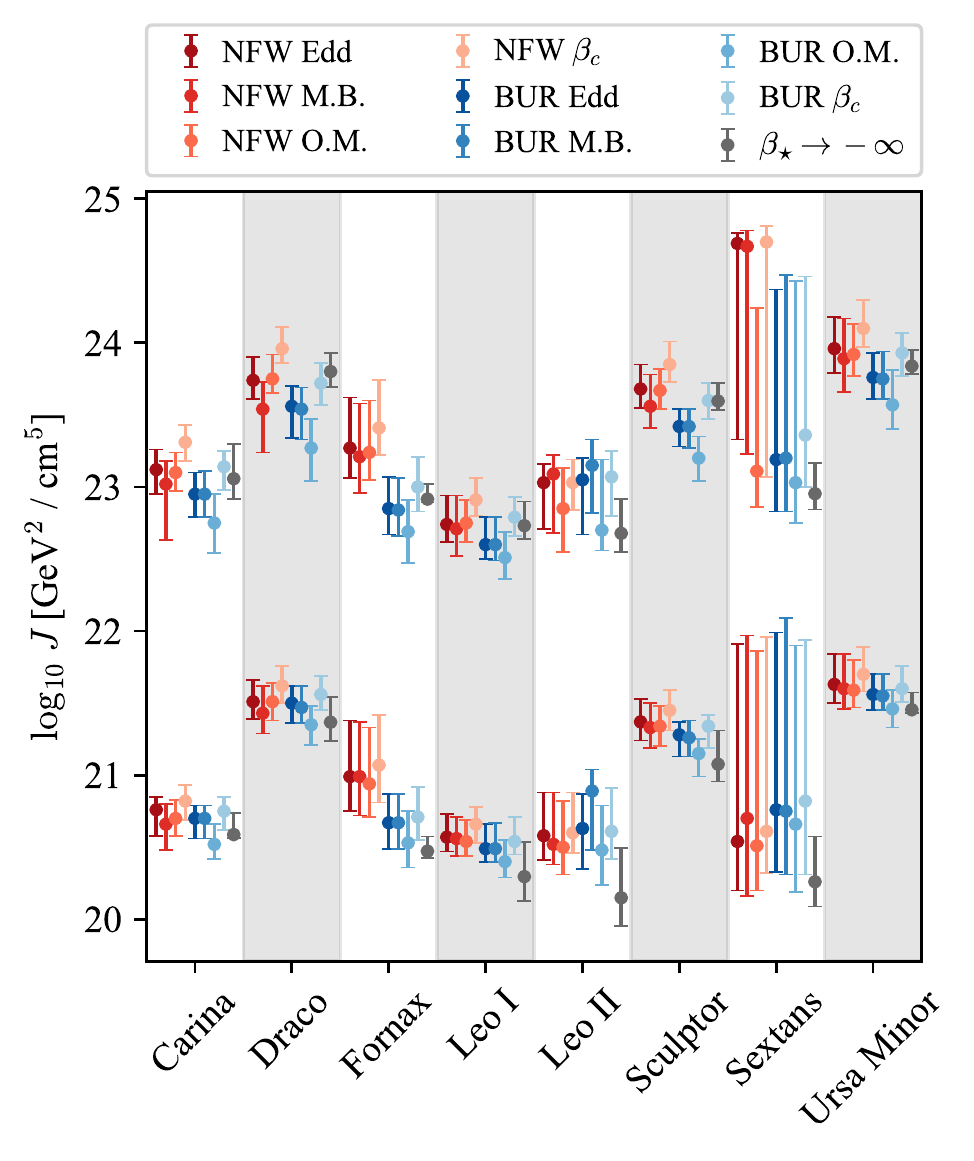}
}
\vspace{-3mm}
\caption{{\emph{Left}: comparison of our results for non-enhanced $J$-factors with previous works. \emph{Right}:~results for Sommerfeld-enhanced $J$-factors for different PSDF under consideration in the Coulomb regime (lower points) and resonant regime (upper points).}}
\label{fig:J_comparison}
\end{figure}

Most importantly, we provide a summary of all our results in figure~\ref{fig:J_comparison}. We display the $J$-factors for the eight classical dSphs in the non-enhanced (left panel), as well as the enhanced regimes (right panel). We report in the figure 68\%  h.p.d. interval extracted from the marginalized posterior distribution. Numerical values displayed in this figure, as well as 95\% h.p.d., are reported in table~\ref{tab:J_factors}. In the left panel of figure~\ref{fig:J_comparison}, we also compare our results in the non-enhanced regime to some recent results in the literature; the good agreement with these --- despite the several sources of differences concerning the parametrization of the stellar surface brightness profile, of the stellar anisotropy profile and the choice DM parametric profile, together with the prior adopted in the corresponding MCMC studies --- comes as further validation of our code, as well as gives a feeling of the impact of other uncertainties in this approach.
The results obtained for the non-parametric profile under the assumption of circular orbits for stars are reported in the plot as 68\% lower limits, since it was shown in~\cite{Ullio-ml-2016kvy} that this case can be used to extrapolate conservative lower limit on (non-enhanced) $J$-factors. From the right panel of figure~\ref{fig:J_comparison} we can read off the following trends: for NFW fits the difference between isotropic and OM model tends to be small, especially for objects that prefer large $r_s$ (in particular, Draco and Carina), which we also adopt as the characteristic scale for the velocity anisotropy, i.e.\ $r_a = r_s$. On the other hand, for the Burkert profile we find stronger dependence on the PSDF anisotropy, which stands out in the same fashion for all the eight objects. In this respect, the prime targets for detection among the eight classical dSphs remain essentially the same as in the non-enhanced case, with slight improvement for Sculptor and Sextans, which we find to have more concentrated DM halos.\footnote{Regarding the peculiar shape of $J_2$ errors for Sextans assuming a NFW profile we need to underline that in this case our results are actually affected by the choice of the inner cut-off radius $r_{\textrm{min}}$ we need to introduce for numerical convergence: a significant number points in the MCMC chain end up at $r_s < r_{\textrm{min}}$, leading to nearly identical $J$-factors; this results in central values lying right at the upper boundary of the 68\% h.p.d. region, with the exception of OM model for which the contribution around $r_s$ is suppressed.} The effect of phase-space modelling turns out to be significant as it is comparable to, or in some cases even exceeds, other uncertainties in the spherical Jeans equation approach and can be summarized as follows: when considering the Sommerfeld enhancement regimes, the OM model may indeed induce up to $\sim 30\%$ decrease in results compared to estimates with Eddington's inversion formula, while $\beta_{\textrm{\tiny DM}}=-1/2$ up to $50\%$ increase. Restraining to the MB approximation implies in general an underestimate of the flux in case of singular or very concentrated profiles. Finally regarding the non-parametric approach of ref.~\cite{Ullio-ml-2016kvy}, 68\% h.p.d.  are shown in the right panel of figure~\ref{fig:J_comparison}, considering the case of Eddington's inversion only.
Note that these do not correspond anymore to the most conservative cases for the extrapolated $J_1$ and $J_2$, as due to the fact that this profile is still singular towards the center and therefore receives a prominent gain in the expected flux from the Sommerfeld effect. A more general analyses would be needed to find the new conservative lower limits in the two enhanced regimes; this is technically and numerically very challenging and beyond the scope of this work.
Another flaw of this density profile is the fact that we cannot actually exploit the OM model to treat radially anisotropic DM configurations, since one finds positive-definite PSDF only for $r_a / r_s \gtrsim 10$,
being essentially equivalent to the Eddington's case.
For this reasons we mostly find the lowest $J$-factors in the enhanced regime for the Burkert profile, which has a flatter central gravitational potential and can be eventually even further suppressed by adopting Osipkov-Merritt's DM orbital anisotropy. As for what regards $\beta_c$ models in context of non-parametric density profile, they are physical and can be computed, however would yield higher $J$-factor, irrelevant for what concern the problem of addressing conservative upper bound on the DM pair annihilation cross-section.

\section{Conclusion}
\label{sec:concl}

Dwarf spheroidal galaxies of the Milky Way occupy a leading role in the indirect search program: in the gamma-ray band they unquestionably offer a golden channel for the detection of the elusive Dark Matter particle.
An accurate prediction of the photon flux due to particle annihilation in the galactic halo of these objects is of great relevance for the community and timely for the progress in the field.

In the recent years, an increasing attention has been devoted to quantify the robustness of such prediction, looping over a variety of uncertainties, mainly of astrophysical origin. The majority of these analyses were restricted to a benchmark case where velocity-dependent effects ---  pertaining to the playground of Particle Physics --- can be neglected. A few recent studies on the subject opened up to a broader investigation, where velocity-dependent contributions from the DM annihilation cross-section have been consistently taken into account in the prediction for the gamma-ray flux, see~\cite{Boddy-ml-2017vpe,Lu-ml-2017jrh,Bergstrom-ml-2017ptx}.

It is important to note that in such analyses one is generally exposed to a large set of systematics, stemming from the lack of any precise information about the galactic phase-space distribution of DM particles.
In this work we have performed the first steps towards a data-based, comprehensive investigation of this issue, inspecting the prediction of the gamma-ray flux from galactic DM yields in relation to the underlying DM phase-space distribution function.  While our approach can be easily applied to any spherical (DM dominated) system, our attention here focussed on the case of the brightest Milky Way dwarf satellites. These are primary targets for DM indirect detection, with a very good sample of measurements on the kinematics of their faint stellar components, which trace the gravitational potential dominated by DM\@. Moreover, the DM velocity probed in dwarf galactic halos falls in the deep non-relativistic regime, where possible long-range interactions --- as described by the well-known Sommerfeld effect~\cite{Hisano-ml-2003ec,Hisano-ml-2004ds,Pospelov-ml-2007mp,ArkaniHamed-ml-2008qn,Feng-ml-2009hw} --- become of significant phenomenological relevance.

We started our investigation by performing a MCMC analysis of the stellar kinematic dataset available for the eight classical dwarf spheroidals.
Within the framework of the spherical Jeans analysis, we estimated the DM halo parameters via Bayesian inference, considering  both the case of cuspy (NFW) and cored (Burkert) halo profiles.
Particular care has been taken for the inclusion of several sources of observational and theoretical uncertainties, some of them overlooked in previous studies, e.g.\ the one concerning the heliocentric distance of the dwarf galaxy.
An optimal description of the stellar kinematics of the classical dwarf spheroidals was obtained for both cuspy and cored  scenarios, see figure~\ref{fig:slos_draco_sculpt}.
Most notably, for these objects we have reassessed the evaluation of the line-of-sight integrated halo density squared, i.e. the so-called $J$-factor. This work presents a state-of-the-art analysis, which gives good agreement with the most recent findings in literature~\cite{Charbonnier-ml-2011ft,Geringer-Sameth-ml-2014yza,Bonnivard-ml-2015xpq}, see figure~\ref{fig:J_comparison}. In the same figure we provide a more conservative estimate of $J$ on the basis of a modelling of the density directly connected to line-of-sight observable quantities, paying once again attention to spectroscopic and photometric error propagation.
For this non-parametric approach, we followed the procedure outlined in detail in~\cite{Ullio-ml-2016kvy,Valli-ml-2016xtu}, and have obtained as a final outcome the lower limits reported as gray arrows in figure~\ref{fig:J_comparison}.

The major novelty of our work consists in a deep study of the generalized notion of the  $J$-factor in order to account for velocity effects stemming from the Sommerfeld enhancement.
In particular, we inspected the crucial interplay of this quantity with the assumption regarding the galactic DM phase-space distribution function.
In practice, we exploited the outcome of our Bayesian fits to indirectly probe four distinct classes of phase-space modelling, without necessarily relying on a vanishing DM anisotropy, the basic assumption of all previous investigations on the subject. Specifically, for what concerns isotropic distribution functions, we have analyzed the case of the Maxwell-Boltzmann approximation
and the limit of ergodic systems via Eddington's inversion formula. Going beyond the standard lore of isotropic phase-space modelling, we have considered the scenario of radially biased DM particle orbits within the so-called Osipkov-Merritt model, see eq.~(\ref{eq:beta_osipkov_meritt}), as well as analyzed tangential-like motion with the study case provided by $\beta_\dm(r) = -1/2$. We discussed the main trends, first considering single particle velocities, then inspecting the impact on $J$-factor via its radial shell contribution.

\enlargethispage*{\baselineskip}

Our main findings in connection with the analyzed observational dataset for Milky Way dwarf spheroidals are reported in the right panel of figure~\ref{fig:J_comparison}: we evaluated the posterior distribution function of the Sommerfeld-enhanced $J$-factors according to the two peculiar limits of phenomenological interest highlighted by eq.~(\ref{eq:Sommnonres})--(\ref{eq:Sommres}). For convenience, the estimated $J$-values with corresponding highest density probability intervals are also reported in table~\ref{tab:J_factors}. In summary, the ordering of prime targets for indirect detection remains essentially the same as in non-enhanced case, however the expected signals are enhanced by several orders of magnitude, depending on the explicit values of particle physics parameters under consideration.
While reassessing the isotropic modeling on the basis of an elaborated data-driven analysis, the key result of the present work
is the evaluation of the $J$-factors for Osipkov-Merritt and tangential-like scenarios: we have obtained a notable boost of $J$ with respect to the corresponding isotropic limit in the case of $\beta_\dm(r) = -1/2$, while we have found a suppression of about the same ratio when the Osipkov-Merritt modelling was assumed instead, with the extent of this suppression actually depending both on the parametrization of the DM profile and the choice of anisotropy scale $r_a$. Since there are no reliable constraints on the velocity anisotropy of DM particles, these results should be interpreted as an increase of the overall uncertainty of $J$-factors for velocity dependent annihilations and taken into account by future analyses.
We ended our investigation by analyzing also the case of non-parametric DM density within the specific assumption of an ergodic system, demanded for this case by the physical condition of a non-negative phase-space distribution function. We reported such scenario in the right panel of figure~\ref{fig:J_comparison} with gray dots, highlighting the fact that in presence of velocity-dependent effects it is no longer representative of a lower-limit for the corresponding~$J$.

In conclusion, our study has provided the first quantitative and data-based attempt in literature to bracket the uncertainty in the prediction of prompt photon emission from DM annihilation stemming from the modeling of the phase-space distribution function in the classical dwarf satellites of the Galaxy. We wish to remark that the results reported in figure~\ref{fig:J_comparison} of this paper may trigger the interest of current experimental facilities such as Fermi-LAT~\cite{bib2009ApJ...697.1071A}, HAWC~\cite{bib2018ApJ...853..154A}, H.E.S.S~\cite{Abramowski-ml-2014tra}, MAGIC~\cite{bib2009AIPC.1166..191S}  and VERITAS~\cite{bib2016chep.confE.446Z}, and may be of further relevance for next-gen ones, such as CTA~\cite{bib2017arXiv170901483M}, $e$-ASTROGAM~\cite{bib2017arXiv171101265D} and AMEGO~\cite{Caputo-ml-2017sjw}. On more general grounds, we believe that the approach outlined in this work may serve as a future baseline for a systematic investigation on the relevance of DM phase-space distribution modelling in the vast realm of indirect searches.

\renewcommand{\floatpagefraction}{.9}

	\begin{sidewaystable}
		\centering
	\def\arraystretch{1.6}
\centerline{	\scalebox{.77}{	\begin{tabular}{|c|c|c|c|c|c|c|c|c|c|}\hline
			dSph & $\log_{10}J_0$ & $\log_{10}J_{1-\textrm{\tiny E}}$ & $\log_{10}J_{1-\textrm{\tiny MB}}$ & $\log_{10}J_{1-\textrm{\tiny OM}}$ & $\log_{10}J_{1-\beta_c}$ & $\log_{10}J_{2-\textrm{\tiny E}}$ & $\log_{10}J_{2-\textrm{\tiny MB}}$ & $\log_{10}J_{2-\textrm{\tiny OM}}$ & $\log_{10}J_{2-\beta_c}$ \\ \hline
			Carina & $17.63^{+0.16(0.29)}_{-0.11(0.23)}$ & $20.76^{+0.09(0.24)}_{-0.18(0.30)}$ & $20.66^{+0.14(0.31)}_{-0.18(0.31)}$ & $20.70^{+0.13(0.26)}_{-0.12(0.24)}$ & $20.82^{+0.11(0.25)}_{-0.13(0.24)}$ & $23.12^{+0.14(0.32)}_{-0.17(0.27)}$ & $23.02^{+0.16(0.35)}_{-0.39(0.58)}$ & $23.10^{+0.14(0.31)}_{-0.13(0.24)}$ & $23.31^{+0.12(0.28)}_{-0.13(0.23)}$ \\
			Draco & $18.67^{+0.12(0.26)}_{-0.15(0.27)}$ & $21.51^{+0.15(0.31)}_{-0.12(0.24)}$ & $21.43^{+0.19(0.36)}_{-0.14(0.28)}$ & $21.51^{+0.13(0.28)}_{-0.13(0.25)}$ & $21.62^{+0.14(0.28)}_{-0.12(0.24)}$ & $23.74^{+0.16(0.34)}_{-0.13(0.24)}$ & $23.54^{+0.19(0.42)}_{-0.30(0.45)}$ & $23.75^{+0.17(0.33)}_{-0.10(0.22)}$ & $23.96^{+0.15(0.30)}_{-0.10(0.21)}$ \\
			Fornax & $17.95^{+0.37(1.19)}_{-0.29(0.43)}$ & $20.99^{+0.39(1.08)}_{-0.24(0.37)}$ & $20.99^{+0.38(1.09)}_{-0.27(0.40)}$ & $20.94^{+0.39(1.05)}_{-0.23(0.37)}$ & $21.07^{+0.35(1.00)}_{-0.26(0.40)}$ & $23.27^{+0.35(0.89)}_{-0.21(0.31)}$ & $23.21^{+0.37(0.95)}_{-0.25(0.37)}$ &
			$23.24^{+0.36(0.85)}_{-0.19(0.34)}$ &
			$23.41^{+0.33(0.82)}_{-0.19(0.29)}$ \\
			Leo I & $17.63^{+0.20(0.43)}_{-0.13(0.24)}$ & $20.57^{+0.16(0.36)}_{-0.10(0.21)}$ & $20.56^{+0.15(0.36)}_{-0.12(0.24)}$ & $20.54^{+0.15(0.34)}_{-0.10(0.20)}$ & $20.66^{+0.12(0.30)}_{-0.13(0.23)}$ & $22.74^{+0.20(0.39)}_{-0.12(0.25)}$ & $22.71^{+0.23(0.46)}_{-0.19(0.46)}$ & $22.75^{+0.16(0.34)}_{-0.13(0.25)}$ &
			$22.91^{+0.15(0.31)}_{-0.11(0.22)}$ \\
			Leo II & $17.65^{+0.21(0.43)}_{-0.31(0.54)}$ & $20.58^{+0.30(0.48)}_{-0.17(0.35)}$ & $20.52^{+0.36(0.54)}_{-0.14(0.33)}$ & $20.50^{+0.32(0.62)}_{-0.19(0.32)}$ & $20.60^{+0.28(0.46)}_{-0.14(0.31)}$ & $23.03^{+0.13(0.24)}_{-0.32(0.57)}$ & $23.09^{+0.13(0.23)}_{-0.41(0.86)}$ &
			$22.85^{+0.28(0.79)}_{-0.30(0.41)}$ &
			$23.03^{+0.16(0.28)}_{-0.19(0.37)}$ \\
			Sculptor & $18.35^{+0.13(0.27)}_{-0.13(0.26)}$ & $21.37^{+0.16(0.34)}_{-0.13(0.28)}$ & $21.33^{+0.17(0.35)}_{-0.14(0.31)}$ & $21.34^{+0.14(0.31)}_{-0.14(0.28)}$ & $21.45^{+0.14(0.30)}_{-0.14(0.27)}$ & $23.68^{+0.17(0.34)}_{-0.13(0.29)}$ &
			$23.56^{+0.22(0.40)}_{-0.15(0.42)}$ &
			$23.67^{+0.15(0.32)}_{-0.13(0.27)}$ &
			$23.85^{+0.16(0.31)}_{-0.12(0.25)}$\\
			Sextans & $17.30^{+1.45(2.49)}_{-0.36(0.38)}$ & $20.54^{+1.37(2.15)}_{-0.34(0.39)}$ & $20.70^{+1.27(2.02)}_{-0.54(0.64)}$ & $20.51^{+1.35(2.17)}_{-0.31(0.38)}$ & $20.61^{+1.35(2.08)}_{-0.29(0.34)}$ & $24.69^{+0.07(0.09)}_{-1.36(1.92)}$ & $24.67^{+0.11(0.18)}_{-1.44(2.20)}$ &
			$23.11^{+1.13(1.78)}_{-0.25(0.39)}$ &
			$24.70^{+0.11(0.33)}_{-1.63(1.70)}$ \\
			Ursa Minor & $18.65^{+0.14(0.34)}_{-0.13(0.26)}$ & $21.63^{+0.21(0.44)}_{-0.13(0.28)}$ & $21.60^{+0.24(0.48)}_{-0.14(0.34)}$ & $21.59^{+0.21(0.44)}_{-0.12(0.24)}$ & $21.70^{+0.19(0.42)}_{-0.12(0.24)}$ &
			$23.96^{+0.22(0.41)}_{-0.17(0.35)}$ & $23.89^{+0.28(0.49)}_{-0.23(0.63)}$ &
			$23.92^{+0.21(0.42)}_{-0.15(0.28)}$ &
			$24.10^{+0.20(0.38)}_{-0.13(0.26)}$\\\hline
		\end{tabular}
}}
		\vspace*{.5 cm}
\centerline{
\scalebox{.77}{
		\begin{tabular}{|c|c|c|c|c|c|c|c|c|c|}\hline
			dSph & $\log_{10}J_0$ & $\log_{10}J_{1-\textrm{\tiny E}}$ & $\log_{10}J_{1-\textrm{\tiny MB}}$ & $\log_{10}J_{1-\textrm{\tiny OM}}$ & $\log_{10}J_{1-\beta_c}$ & $\log_{10}J_{2-\textrm{\tiny E}}$ & $\log_{10}J_{2-\textrm{\tiny MB}}$ & $\log_{10}J_{2-\textrm{\tiny OM}}$ & $\log_{10}J_{2-\beta_c}$ \\ \hline
			Carina & $17.57^{+0.18(0.40)}_{-0.13(0.25)}$ & $20.70^{+0.09(0.23)}_{-0.14(0.27)}$ & $20.70^{+0.09(0.22)}_{-0.14(0.27)}$ & $20.52^{+0.14(0.28)}_{-0.10(0.24)}$ & $20.75^{+0.10(0.23)}_{-0.13(0.24)}$ & $22.95^{+0.15(0.35)}_{-0.16(0.40)}$ & $22.95^{+0.16(0.36)}_{-0.16(0.40)}$ & $22.75^{+0.20(0.42)}_{-0.21(0.52)}$ & $23.14^{+0.11(0.28)}_{-0.16(0.32)}$ \\
			Draco & $18.67^{+0.16(0.32)}_{-0.13(0.25)}$ & $21.50^{+0.12(0.26)}_{-0.14(0.26)}$ & $21.47^{+0.15(0.29)}_{-0.11(0.22)}$ & $21.35^{+0.13(0.27)}_{-0.14(0.27)}$ & $21.56^{+0.13(0.26)}_{-0.11(0.22)}$ & $23.56^{+0.14(0.31)}_{-0.22(0.46)}$ & $23.54^{+0.15(0.32)}_{-0.21(0.45)}$ & $23.27^{+0.20(0.41)}_{-0.23(0.48)}$ & $23.72^{+0.14(0.30)}_{-0.15(0.29)}$ \\
			Fornax & $17.67^{+0.21(0.80)}_{-0.14(0.30)}$ & $20.67^{+0.20(0.78)}_{-0.18(0.32)}$ & $20.67^{+0.20(0.78)}_{-0.18(0.32)}$ & $20.53^{+0.22(0.81)}_{-0.17(0.33)}$ & $20.71^{+0.21(0.77)}_{-0.16(0.32)}$ & $22.85^{+0.22(0.76)}_{-0.18(0.35)}$ & $22.84^{+0.22(0.76)}_{-0.18(0.36)}$ &
			$22.69^{+0.22(0.86)}_{-0.22(0.40)}$ &
			$23.00^{+0.21(0.73)}_{-0.17(0.33)}$ \\
			Leo I & $17.59^{+0.16(0.38)}_{-0.14(0.26)}$ & $20.49^{+0.17(0.37)}_{-0.09(0.21)}$ & $20.49^{+0.18(0.37)}_{-0.09(0.21)}$ & $20.40^{+0.15(0.37)}_{-0.11(0.22)}$ & $20.54^{+0.17(0.36)}_{-0.09(0.21)}$ & $22.60^{+0.19(0.47)}_{-0.10(0.21)}$ & $22.60^{+0.19(0.46)}_{-0.11(0.21)}$ &  $22.51^{+0.18(0.50)}_{-0.15(0.27)}$ &
			$22.79^{+0.14(0.38)}_{-0.13(0.23)}$ \\
			Leo II & $17.55^{+0.24(0.47)}_{-0.30(0.51)}$ & $20.63^{+0.24(0.43)}_{-0.28(0.47)}$ & $20.89^{+0.15(0.26)}_{-0.41(0.70)}$ & $20.48^{+0.31(0.53)}_{-0.24(0.43)}$ & $20.61^{+0.30(0.47)}_{-0.19(0.40)}$ & $23.05^{+0.15(0.27)}_{-0.38(0.64)}$ & $23.15^{+0.18(0.24)}_{-0.33(0.69)}$ &
			$22.70^{+0.49(0.66)}_{-0.14(0.41)}$ &
			$23.07^{+0.18(0.31)}_{-0.27(0.50)}$ \\
			Sculptor & $18.30^{+0.12(0.26)}_{-0.12(0.24)}$ & $21.28^{+0.09(0.23)}_{-0.15(0.27)}$ & $21.26^{+0.12(0.24)}_{-0.13(0.25)}$ & $21.15^{+0.10(0.23)}_{-0.16(0.29)}$ & $21.34^{+0.08(0.21)}_{-0.15(0.27)}$ & $23.42^{+0.12(0.27)}_{-0.14(0.30)}$ &
			$23.42^{+0.12(0.26)}_{-0.15(0.30)}$ &
			$23.20^{+0.15(0.33)}_{-0.16(0.33)}$ &
			$23.60^{+0.12(0.25)}_{-0.13(0.28)}$\\
			Sextans & $17.41^{+1.26(2.27)}_{-0.44(0.52)}$ & $20.76^{+1.23(1.97)}_{-0.43(0.69)}$ & $20.75^{+1.34(2.13)}_{-0.44(0.67)}$ & $20.66^{+1.24(2.03)}_{-0.47(0.77)}$ & $20.82^{+1.12(1.89)}_{-0.51(0.68)}$ & $23.19^{+1.18(1.70)}_{-0.36(0.81)}$ & $23.20^{+1.27(1.79)}_{-0.37(0.80)}$ &
			$23.03^{+1.40(1.91)}_{-0.28(0.95)}$ &
			$23.36^{+1.10(1.54)}_{-0.36(0.72)}$ \\
			Ursa Minor & $18.58^{+0.14(0.30)}_{-0.11(0.23)}$ & $21.56^{+0.14(0.33)}_{-0.11(0.22)}$ & $21.55^{+0.15(0.33)}_{-0.10(0.21)}$ & $21.46^{+0.13(0.32)}_{-0.13(0.26)}$ & $21.60^{+0.16(0.34)}_{-0.09(0.20)}$ &
			$23.76^{+0.17(0.40)}_{-0.15(0.31)}$ & $23.75^{+0.19(0.41)}_{-0.14(0.31)}$ &
			$23.57^{+0.24(0.50)}_{-0.17(0.37)}$ &
			$23.93^{+0.14(0.37)}_{-0.16(0.28)}$\\\hline
		\end{tabular}}}
		\caption{{J-factors with $1\sigma$ ($2\sigma$) errors for the 8 dSph with NFW and Burkert density profile.}}
		\label{tab:J_factors}
	\end{sidewaystable}

\acknowledgments

M.P. and P.U. acknowledge partial support from the European Union's
Horizon 2020 research and innovation programme under the Marie
Sk{\l}odowska-Curie grant agreement No 690575 and from the European
Union's Horizon 2020 research and innovation programme under the Marie
Sklodowska-Curie grant agreement No 674896. M.V.  thanks organizers and participants of the stimulating workshop ``WIMPs vs non-WIMPs in dwarf spheroidal galaxies'' (University of Turin).

\appendix

\section{Details on the computations of the present work}

\subsection{Boundary conditions in Eddington's formula} \label{app:boundary}

For ergodic spherical systems, Eddington's formula, eq.~\eqref{eq:Eddington}, allows to retrieve the phase space distribution function
$f(\mathcal{E})$ corresponding to a given radial density profile $\rho(r)$. The derivation of the formula starts with
noticing that at any radius $r$, the integral over velocities of $f$ (giving $\rho$) can be rewritten in terms of
the relative potential and energy, respectively, $\Psi(r) = -\Phi(r) + \Phi_b$ and $\mathcal{E}=\Psi(r)-|\vec{v}|^2/2$:
\begin{equation}
   \label{eq:PSint}
   \rho(r) = \int d^3\vec{v} \,f(\mathcal{E}) =  4\pi \int_0^\Psi  d\mathcal{E} \; \sqrt{2(\Psi(r) - \mathcal{E})} \; f(\mathcal{E})\,.
\end{equation}
Exploiting the fact that for any spherical system $\Psi$ is a monotonic function of $r$, one can treat $\Psi$  as dependent
variable to show that the derivative of $\rho$ with respect to $\Psi$ and $f(\mathcal{E})$ are in one-to-one correspondance
via an Abel integral equation:
\begin{equation}
   \frac{d\rho}{d\Psi} = 2 \sqrt{2} \pi \int_0^\Psi  d\mathcal{E} \; \frac{f(\mathcal{E})}{\sqrt{\Psi - \mathcal{E}}}\,;
\end{equation}
this can be inverted to find the expression for the distribution function given in Eddington's formula.
This short recap was to remark that a given ergodic $f(\mathcal{E})$ is indeed selected for \emph{i)} a given $d\rho/d\Psi$, plus
\emph{ii)} the boundary condition $\rho(r_b)=0$ at $\Psi(r_b)=0$ (as follows from the expression in eq.~\eqref{eq:PSint}).
Usually one takes the limit of an isolated system of infinite size, extrapolating $r_b \rightarrow \infty$ (and hence
$\Psi(r_b) \rightarrow 0$ corresponding to $\Phi_b=0$); in the case at hand however, since we are considering fairly small
systems, we need to deal with a finite $r_b$  and implement a different procedure.

Our starting point is a parametric form for $\rho_{\textrm{\tiny DM}}(r)$, whose structure parameters, the scale radius and
a reference density, are fitted against dynamical data. Still, while dynamical informations constrain mainly the inner portion
of the profile, say the portion within the half-light radius of the object, the fit is extrapolated outside this region and up to
some shell at which the density profile is truncated, accounting for the fact that dwarf satellites are well within the potential
well of the Galaxy and tidal effects reshape its outskirt. Among possible ways to account for such truncation, we discuss
here two.

As a first possibility, one may simply introduce a sharp cutoff in the density profile, keeping the parametric form unchanged
up to the boundary $r_b$ and just imposing $\rho_{\textrm{\tiny DM}}(r>r_b)=0$. This procedure gives, up to $r=r_b$, the
same $d\rho/d\Psi$ as in the case $r_b \rightarrow \infty$, however it is inconsistent with the boundary condition on $\rho$:
the phase space distribution function $f(\mathcal{E})$, as retrieved from Eddington's formula, does not trace $\rho$ but
actually:
\begin{equation}
   \label{eq:norho}
   \int d^3\vec{v} \,f(\mathcal{E}) =  \rho(r) - \rho(r_b)\,.
\end{equation}
E.g., for the Hernquist profile~\cite{bib1990ApJ...356..359H}:
\begin{equation}
   \label{eq:Hernrho}
   \rho_H(r) = \frac{\rho_s}{(r/r_s)\,(1+r/r_s)^3}
\end{equation}
with sharp truncation at $r_b$, one has:
\begin{equation}
   \label{eq:HernPsi}
   \Psi_H(r) = 2 \pi G_N \rho_s r_s^2 \left(\frac{1}{1+r/r_s} - \frac{1}{1+r_b/r_s}\right) \equiv \Psi_\infty(r) -  \Psi_b
\end{equation}
where $\Psi_\infty$ and $\Psi_b$ have been defined correspondingly to the two terms in parenthesis, and $\Psi_\infty$
refers to the result for $r_b \rightarrow \infty$. As in the case without the truncation, $\Psi_H(r)$ can be inverted and the
integral in Eddington's formula can be performed analytically; one finds:
\begin{eqnarray}
   \label{eq:Hernf}
    f_H(\mathcal{E}) & = & \frac{\rho_s}{ 4 \sqrt{2}\,\pi^2 C^{3/2}}\nonumber\\
&&\times \Bigg\{ \sqrt{\widetilde{\mathcal{E}}} \left[
    \frac{3}{(1-\widetilde{\Psi}_b)\,(1-\widetilde{\Psi}_b-\widetilde{\mathcal{E}})^2}
    +\frac{2}{(1-\widetilde{\Psi}_b)^2\,(1-\widetilde{\Psi}_b-\widetilde{\mathcal{E}})}
    -8 (1+3\widetilde{\Psi}_b+2\widetilde{\mathcal{E}}) \right]    \nonumber \\
    &&
    +\frac{3}{(1-\widetilde{\Psi}_b-\widetilde{\mathcal{E}})^{5/2}}
    \arctan\Bigg( \frac{\sqrt{\widetilde{\mathcal{E}}}}{\sqrt{1-\widetilde{\Psi}_b-\widetilde{\mathcal{E}}}}\Bigg)
    +\frac{2}{\sqrt{\widetilde{\mathcal{E}}}} \frac{\widetilde{\Psi}_b^3}{(1-\widetilde{\Psi}_b)^2}  (4-3\widetilde{\Psi}_b) \Bigg\}
\end{eqnarray}
with $C\equiv 2 \pi G_N \rho_s r_s^2$ and $\widetilde{X} \equiv X/C$ for any of the quantities $X$ above.
Taking the limit $\widetilde{\Psi}_b\rightarrow 0$, $f_H(\mathcal{E})$ correctly reduces to the expression
for the distribution function given in~\cite{bib1990ApJ...356..359H} when assuming $r_b \rightarrow \infty$. Integrating over
velocities, one indeed finds:
\begin{equation}
   \label{eq:Hernrhoder}
   \int d^3\vec{v} f_H(\mathcal{E}) =  \rho_s \left(\frac{\widetilde{\Psi}_\infty^4}{1-\widetilde{\Psi}_\infty}-\frac{\widetilde{\Psi}_b^4}{1-\widetilde{\Psi}_b}\right)
   =  \rho_H(r) -  \rho_H(r_b)\,.
\end{equation}
To fill in the density gap one could consider to add to $f(\mathcal{E})$ an extra term.
One sample choice could be:
\begin{equation}
   \label{eq:extraf}
   f_b(\mathcal{E},|\vec{v}|) =  \frac{\rho(r_b)}{4\pi} \frac{1}{|\vec{v}|^2\,\sqrt{2 \mathcal{E}+|\vec{v}|^2}}\,,
\end{equation}
While this correction is crucial for accurate density reconstruction, it has a smaller impact on $J$-factors, which are dominated by the central contribution where the relative shift in density is in general small, i.e.\ $\rho^2_\dm(r \sim r_s) \gg \rho^2_\dm(r_b)$ for typical truncation radius $r_b$. This remains true also in the velocity dependent regimes, since $f(\mathcal{E})$ is even more sharply peaked then the proposed $f_b(\mathcal{E}, |\vec{v}|)$ in the $\mathcal{E} \rightarrow \Psi(0)$ (which is the equivalent of $\vec{v} \rightarrow 0$) limit. We checked this also numerically and indeed found even smaller effect on $J$-factors then in the velocity-independent case.

As a second possibility to modify the density and impose $\rho_{\textrm{\tiny DM}}(r>r_b)=0$, one can
consider to introduce a smoothing function $s(r)$ and make the replacement:
\begin{equation}
   \label{eq:rhosm}
   \forall \; r_{sm} \le r \le r_b  \quad \quad \rho(r) \rightarrow \rho(r) \cdot s(r)\,,
\end{equation}
taking care that the new density profile and its derivative are continuous in $r_{sm}$ and $r_b$, hence imposing:
\begin{equation}
   \label{eq:ssmcond}
   s(r=r_{sm})=1, \quad
   s(r=r_{b})=0, \quad
   \frac{ds}{dr}(r=r_{sm})=1 \quad \& \quad
   \frac{ds}{dr}(r=r_{b})=0.
\end{equation}
The lowest order polynomial function satisfying these conditions is:
\begin{equation}
   \label{eq:ssmfunc}
   s(r)= 3\,\frac{(r_b-r)^2}{(r_b-r_{sm})^2}-2\,\frac{(r_b-r)^3}{(r_b-r_{sm})^3}\,.
\end{equation}
While this approach in principle reduces to the sharp cutoff case when $r_{sm} \rightarrow r_b$ it is not in general
possible to take this limit smoothly: there is no guarantee that Eddington's formula provides a physical result, namely
that it satisfies the physical requirement that $f(\mathcal{E})$ is positive definite; this is indeed the case for any of
the parametric profiles considered in this analysis, but it can be violated when the profile is modified according
to the prescription in eq.~\eqref{eq:rhosm} if the smoothing function makes the profile change too rapidly. For a given
profile and a given $r_b$, one can find numerically the largest possible $r_{sm}$ for which $f(\mathcal{E})$ is physical. For both NFW and Burket profiles we find as a upper limit on the smoothing scale $r_{sm}  \approx 0.67 r_b$, nearly independent of $r_b$ as long as $r_b \gg r_s$. The situation is slightly worse for our non-parametric profile, obtained from the Jeans inversion method, for which $f(\mathcal{E})$ is positive definite as long as $r_{sm} / r_b \lessapprox 0.58$. Smooth truncation has even smaller effect on $J$-factors as the central part within $r_{sm}$ remains unmodified. Apart from this, smooth truncation has also advantage in being easier to treat numerically and does not need an ad-hoc correction term. We verified that in general other smoothing functions, such as a larger order polynomials, require even smaller $r_{sm}$ and hence introduce larger impact on the $J$-factors.

The discussion in case of the Osipkov-Merritt models is specular; the only extra ingredient is that also the radial
anisotropy factor $(1 + r^2 / r_a^2)$ may drive the Abel inversion formula to an unphysical $f(\mathcal{Q})$. One finds
that in general $r_a$ cannot be extrapolated below a given fraction of the scale radius $r_s$, depending on the
parametric form of the profile, as well as on $r_b$ and eventually $r_{sm}$ (this is true also when extrapolating
$r_b \rightarrow \infty$). For instance, for a NFW profile we found the minimum $r_a / r_s \approx 0.35$ while for a Burket profile the situation is slightly worse, with the limit at $r_a / r_s \approx 0.75$; these values are again roughly constant as long as $r_s \ll r_{sm} \ll r_b$. As a final remark we note that our non-parametric density profile is essentially incompatible with Osipkov-Merritt model truncated by $s(r)$. We found non-negative $f(\mathcal{Q})$ only for $r_a / r_s \gtrsim 10$, which practically coincides with the isotropic case.

\subsection{Integrals over velocities of the DM distribution functions}

As already discussed in section~\ref{sec:J_factors}, when the annihilation probability depends on the relative velocity of the DM particles
in a pair, the $J$-factor integrand for spherically symmetric DM profiles contains the radially
dependent rescaling function $\langle S (v_{\textrm{rel}}) \rangle(r)$ introduced in eq.~\eqref{eq:J-integrand}. In most cases it cannot be computed analytically; it is then essential to implement an efficient
numerical calculation, taking into account what variable(s) $f_\dm$ is effectively dependent on.

For ergodic cases, since each distribution function is in form $f_\dm(\mathcal{E}_i)$ with
$\mathcal{E}_i= \mathcal{E}_i(r,|\vec{v}_i|)$, one possible choice is to perform the integral by changing
variables from the velocity of the two particles $(\vec{v}_1,\vec{v}_2)$ to the center of mass velocity and
relative velocity $(\vec{v}_{\rm cm},\vec{v}_{\rm rel})$ (namely $\vec{v}_{\rm cm} = (\vec{v}_1+\vec{v}_2)/2$ and
$\vec{v}_{\rm rel} =\vec{v}_1-\vec{v}_2$). As detailed, e.g., in~\cite{Ferrer-ml-2013cla}, this choice leads to a three
dimensional integral in $|\vec{v}_{\rm cm}|$, $|\vec{v}_{\rm rel}|$ and the angle between $\vec{v}_{\rm cm}$ and
$\vec{v}_{\rm rel}$. In case of MB velocity distributions, see eq.~(\ref{eq:MB_PSDF}), this is particularly convenient since the integral
over the angle and $|\vec{v}_{\rm cm}|$ can be performed analytically and only a 1-dimensional numerical integral
remains; except if you need to track the dependence of result on $|\vec{v}_{\rm rel}|$, in all other ergodic cases
these there is actually no gain with respect to keeping the initial variable choice and reduce to a numerical
computation in $|\vec{v}_1|$, $|\vec{v}_2|$ and the angle between $\vec{v}_1$ and $\vec{v}_2$ (which is
essentially what we do in our numerical implementation, we only replace $|\vec{v}_i|$ with $\mathcal{E}_i$).

For the general case, in which the distribution function depends on both $\mathcal{E}_i$ and the modulus of the
angular momentum $L_i$, the computation of $\langle S (v_{\textrm{rel}}) \rangle(r)$ involves 5 numerical integrals at any
given value of $r$, definitely too demanding for large parameter space scans. In the two cases considered
in this paper, the Osipkov-Merritt models and models with factorizable $L_i$ dependence and constant anisotropy,
the numerical evaluation can, however, be simplified. The natural choice is to perform the integral in spherical coordinates for
the velocity space of each of the two particles, choosing the polar angle $\eta_i$ as the angle between $\vec{v}_i$ and
$\vec{r}$  (and hence with $L_i = r\,|\vec{v}_i| S_i$; here and below we have shortened the notation introducing
$S_i \equiv \sin \eta_i$ and $C_i \equiv \cos \eta_i$). It is then convenient to replace the integral over the azimuthal
angles $\phi_1$ and  $\phi_2$ with those on $\phi_+ \equiv (\phi_1 + \phi_2)/2$ and $\phi_{\rm rel} \equiv \phi_1 - \phi_2$, given that there is no dependence on the azimuthal angles in $f_\dm$, and $|\vec{v}_{\rm rel}|$ depends
only~on~$\phi_{\rm rel}$:
\begin{equation}
\label{eq:vrel}
|\vec{v}_{\rm rel}| = [|\vec{v}_1|^2 + |\vec{v}_2|^2 - 2  |\vec{v}_1|  |\vec{v}_2| (S_1 S_2 \cos \phi_{\rm rel} + C_1 C_2)]^{1/2}\,.
\end{equation}
The integration over $\phi_+$ is then trivial, giving a factor of $2 \pi$. Of the remaining 5 integrals,
we can consider first those 3 with integrand that --- in our reference cases --- does not (even implicitly)
depend on any of the parameters for the Bayesian fits of the density profiles. When performing a scan,
this part of the numerical calculation can be precomputed and stored as a tabulated function, to be linked
for every point in the parameter space. The general structure takes the form:
\begin{equation}
\langle S (v_{\textrm{rel}}) \rangle(r) = \frac{2\pi}{\rho_\dm^2(r)}  \int_{0}^{\Psi(r)} dX_1 \, g_\dm(X_1)
\int_{0}^{\Psi(r)} dX_2 \, g_\dm(X_2) \, \mathcal{S}(r,X_1,X_2)
\end{equation}
with the function to be tabulated:
\begin{equation}
\label{eq:Save}
\mathcal{S}(r,X_1,X_2) =
\int_{-1}^{+1} dC_1 \frac{|\vec{v}_1|^2}{L_1^{2\beta_c}} \left|\frac{\partial |\vec{v}_1|}{\partial X_1}\right|
\int_{-1}^{+1} dC_2 \frac{|\vec{v}_2|^2}{L_2^{2\beta_c}} \left|\frac{\partial |\vec{v}_2|}{\partial X_2}\right|
\int_0^{2\pi} d\phi_r \,S(|\vec{v}_{\rm rel}|) \,.
\end{equation}
Here we have defined, in case of the Osipkov-Merritt models:
\begin{equation}
X_i \equiv Q_i=\Psi(r) - (1+ r^2/r_a^2 \, S_i^2) \,|\vec{v}_i|^2/2\,, \quad \quad
g_\dm(X_i) \equiv f_{\dm,\, {\textrm{\tiny OM}}}(Q_i)  \,, \quad \quad\beta_c \equiv 0\,,
\end{equation}
while, for the models with constant anisotropy $\beta_c\ne 0$:
\begin{equation}
X_i \equiv \mathcal{E}_i=\Psi(r) - \,|\vec{v}_i|^2/2\,, \quad \quad
g_\dm(X_i) \equiv f_{\dm,\, \beta_c}(\mathcal{E}_i,L_i=L_0)\,.
\end{equation}
The picture simplifies further when $S(|\vec{v}_{\rm rel}|)$ describes the enhancement in the flux due to
the Sommerfeld effect, and in case one is interested only at the Coulomb or the resonant regimes in the
Hulten's potential, with scalings, respectively, $S(|\vec{v}_{\rm rel}|) \propto 1/|\vec{v}_{\rm rel}|$ and
$S(|\vec{v}_{\rm rel}|) \propto 1/|\vec{v}_{\rm rel}|^2$. Recasting the expression in eq.~(\ref{eq:vrel}) above
in the form $|\vec{v}_{\rm rel}| \equiv (A - B \cos \phi_{\rm rel})^{1/2}$, one finds:
\begin{equation}
\int_{0}^{2 \pi} \phi_{\rm rel} \, |\vec{v}_{\rm rel}|^{-\alpha} =
\left\{\begin{array}{lr}\displaystyle
        \frac{4}{\sqrt{A - B}} \; \mathrm{K} \left(-\frac{2B}{A-B} \right) & \quad \text{for } \alpha = 1\\[5mm]\displaystyle
        \frac{2 \pi}{\sqrt{A^2 - B^2}} & \quad \text{for }  \alpha = 2
        \end{array} \right.
\end{equation}
where $\mathrm{K}(m)$ is the complete elliptic integral of the first kind:
\begin{equation}
  \mathrm{K}(m) = \int_0^{\frac{\pi}{2}} \frac{dx}{\sqrt{1-m\, \sin^2x}}\;,
\end{equation}
a function which can be very efficiently approximated through its relation to arithmetic-geometric mean, see e.g.~\cite{bib1965hmfw.book.....A}. The integrals on $C_1$ and $C_2$ in eq.~(\ref{eq:Save})
have still to be performed numerically, however the dependence on the external variables partially factorizes:
introducing $\bar{X}_i \equiv X_i/\Psi$, being the velocity of the two particles in the form
$|\vec{v}_i|= \sqrt{\Psi}  \sqrt{1-\bar{X}_i}\,F(r,\eta_i)$, we redefine:
\begin{equation}
  \sqrt{1 - \bar{X}_1} \equiv \mathcal{R} \cos \vartheta\quad {\rm and} \quad   \sqrt{1 - \bar{X}_2} \equiv \mathcal{R} \sin \vartheta \,,
\end{equation}
i.e.:
\begin{equation}
\mathcal{R} = \sqrt{2 - \bar{X}_1  - \bar{X}_2} \quad {\rm with} \quad 0\le \mathcal{R} \le \sqrt{2}\,,
\quad {\rm and} \quad
\tan \vartheta = \sqrt{\frac{1 - \bar{X}_2}{1 - \bar{X}_1}}  \quad {\rm with} \quad 0\le \vartheta \le \frac{\pi}{2}\,.
\end{equation}
It follows that one can extract the scalings in $\sqrt{\Psi}$ and $\mathcal{R}$ in eq.~(\ref{eq:Save}):
\begin{equation}
\label{eq:barSave}
\mathcal{S}(r,X_1,X_2) =  \left(\sqrt{\Psi}\,\mathcal{R}\right)^{2-\alpha-4\beta_c}
\bar{\mathcal{S}}(r,\vartheta)\,,
\end{equation}
where the scaling with $r$ of the new function we introduced  goes simply as $r^{-4\beta_c}$ for the case of
constant anisotropy models, and is very close to a power law even for Osipkov-Merritt models when $r\gg r_a$.
Finally, using symmetries of the integrand function one can show that:
\begin{equation}
\bar{\mathcal{S}} = \bar{\mathcal{S}}(r,{\pi}/{2}-\vartheta) \quad {\rm for}   \quad {\pi}/{4}\le \vartheta \le {\pi}/{2}\,,
\end{equation}
and that the numerical integrals on $C_1$ and $C_2$ can be both reduces to the interval $[0,1]$.

\section{\texorpdfstring{\boldmath $J$}{J}-factor scaling relation} \label{app:J-factor_scaling}

\begin{figure}
	\centerline{
	\includegraphics[scale=0.76]{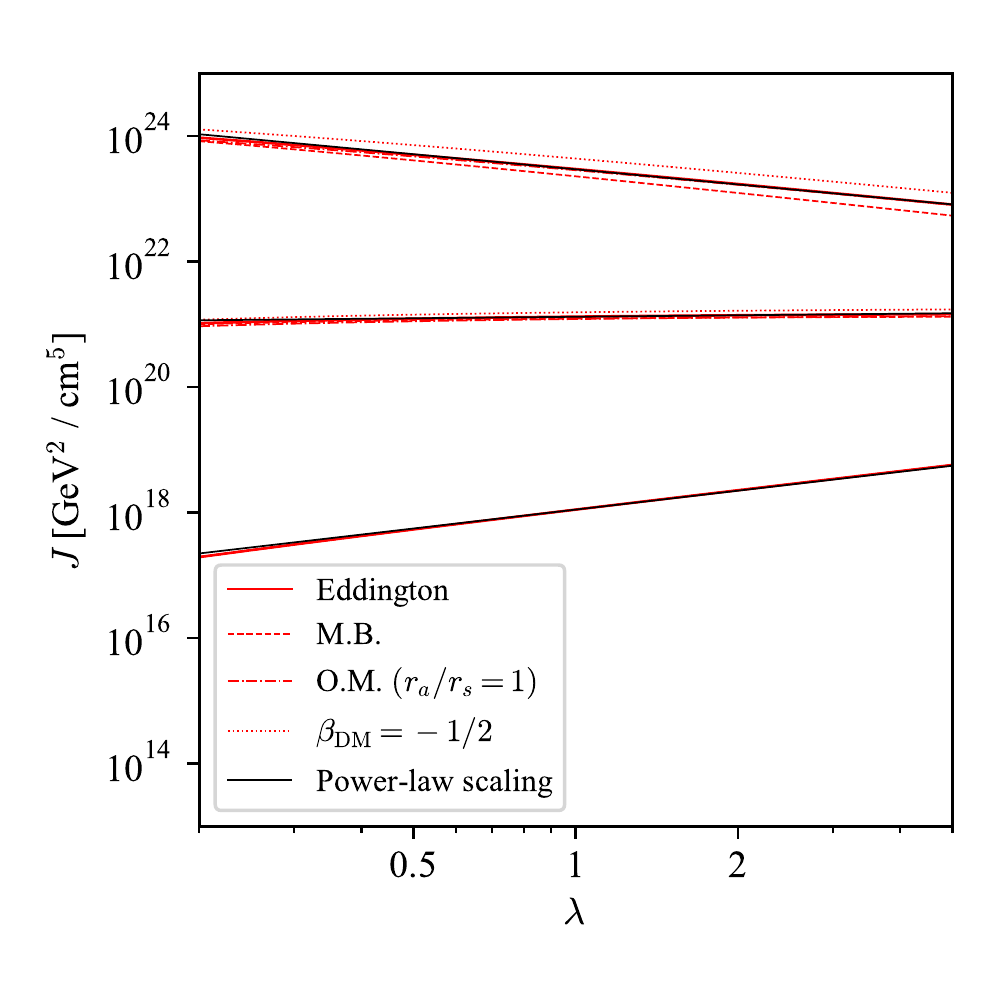}
	\includegraphics[scale=0.76]{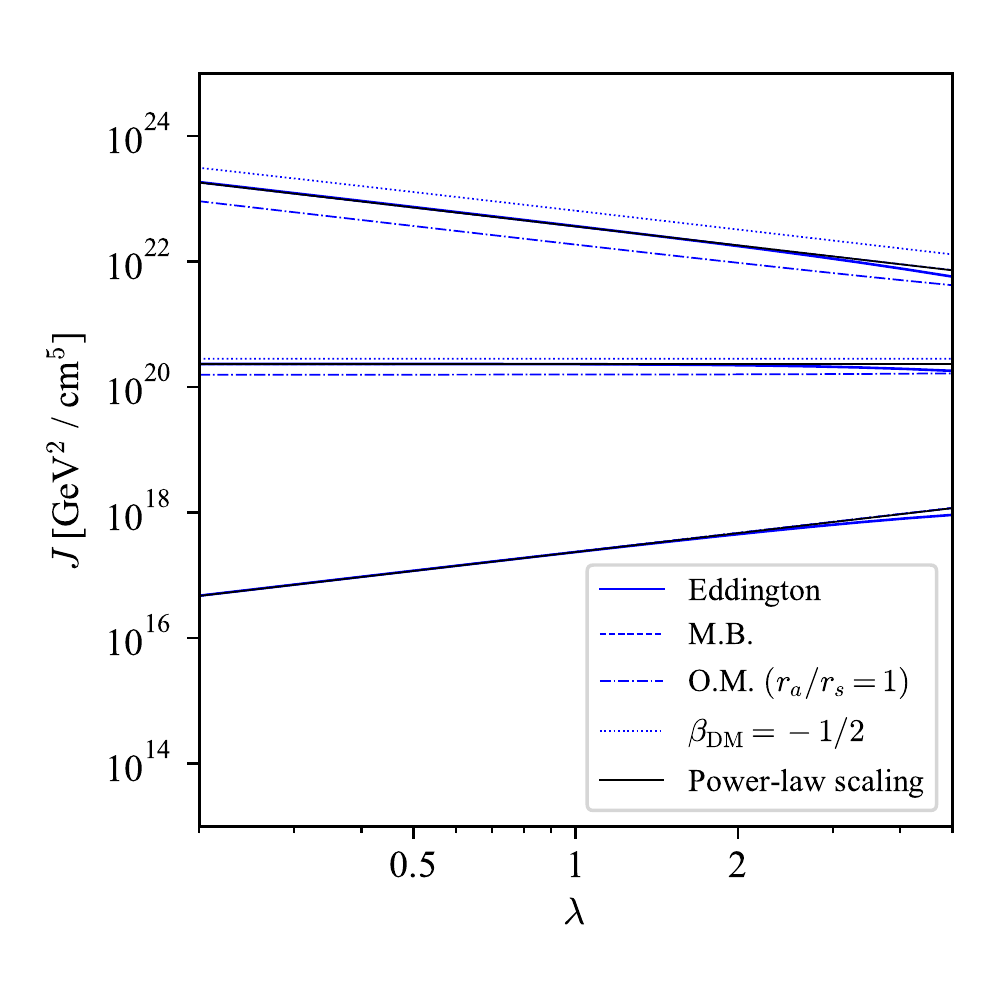}
}
\vspace{-6mm}
	\caption{{$J$-factor scaling under change of physical length $r \rightarrow \lambda r$ for NFW and Burkert profiles with nominal values of $r_s = 1$ kpc, $D = 100$ kpc and $\rho_s = 1$\,GeV/cm$^3$.}}
	\label{fig:J_scaling}
\end{figure}

Precise evaluation of $J$-factors turns out to be too computationally demanding to run it for all the samples produced by the MCMC walkers, even when using the shortcut discussed in previous section. Therefore we resort to scaling relations which allow us to extrapolate the values to arbitrary halo parameters without the need of recomputing the nested numerical integrals in eq.~\eqref{eq:j_factor}. More precisely, it is possible to obtain exact scaling relations for a change in the DM scale density $\rho_s \rightarrow \mu \rho_s$ and physical length scale $r \rightarrow \lambda r$, which also applies to all distance parameters, e.g.\ $r_s$ and $D$. For all spherically symmetric density profiles of the form $\rho_{\textrm{\tiny DM}}(r; \rho_s, r_s) = \rho_s \cdot f(r / r_s)$, where $f(x)$ is an arbitrary function, it is possible to show the following relation:
\begin{align}
\label{eqn:J_scaling}
J(r_s / D \; ; \; \mu \rho_s, \lambda r_s, \lambda D) = J(r_s / D \; ; \; \rho_s, r_s, D) \times
\begin{cases}
\mu^2 \lambda \ \ \ \ \textrm{for} \ \, \xi \gg 1 \,, \\[2mm]
\mu^{3/2}  \ \ \ \, \textrm{for} \ \, \xi \ll 1 \,, \\[2mm]
\mu/\lambda \ \ \ \ \textrm{for} \ \, \xi = 6/\left(\pi^2 n^2\right) \ .
\end{cases}
\end{align}
While the effect of varying $\rho_s$ is quite clear, the scaling with $\lambda$ is somewhat less expected and is in fact broken by the radial cut-offs, which we kept constant. We numerically checked the deviations from the above relation and found excellent agreement except for resonantly enhanced NFW profile, where we found scaling exponent of $\sim -0.8$ instead of $-1$. This is a consequence of using static inner cut-off, which removes an increasingly significant amount of the DM cusp for decreasing values of $\lambda$. We present the comparison of power-law scalings with the numerical results in figure~\ref{fig:J_scaling}. Some deviations also arise at small/large values of $\lambda$ where the effect of cut-offs again becomes noticeable, however when considering dSphs one mostly deals with $\lambda \sim \mathcal{O}(1)$. Using this shortcut the $J$-factors can be thought of as only a function of $r_s / D$, for which however we found no analytical form. Instead we interpolated $J(r_s/D)$ which, together with scaling relations, allowed us to compute accurate $J$-factor posterior distributions from the entire MCMC sample.

\end{document}